\documentclass[12pt]{article}

\usepackage{jheppub} 
\usepackage{float}
\usepackage{amsmath,bbm,blkarray,bm}
\usepackage[vcentermath]{youngtab}
\usepackage{slashed}
\usepackage{bbold}
\usepackage{booktabs}
\usepackage{multirow}
\usepackage{tikz}
\usepackage{changepage}
\usepackage{makecell}
\usepackage{tikz-feynman}
\usepackage{subcaption}
\usepackage{color}
\usepackage[numbers,sort&compress]{natbib}
\usepackage[T1]{fontenc} 

\usepackage{adjustbox}
\usepackage{cancel}

\newcommand{\op}[3]{#1^{#2}_{#3}}

\newcommand{\cop}[3]{\widetilde #1^{#2}_{#3}}

\newcommand{\1}{\mathbb{1}}

\allowdisplaybreaks

\title{One-loop Fierz Transformations}

\author[a]{Jason~Aebischer,}
\author[a]{Marko Pesut}

\affiliation[a]{Physik-Institut, Universit\"at Z\"urich, CH-8057 Z\"urich, Switzerland}

\emailAdd{jason.aebischer@physik.uzh.ch}
\emailAdd{marko.pesut@physik.uzh.ch}

\abstract{
Fierz transformations for four-fermion operators are generalized to the one-loop level. A general renormalization scheme is used to compute QCD and QED corrections to the tree-level relations, which result from Fierz-evanescent operators. The results can be used to perform general one-loop basis transformations involving four-fermi and evanescent operators. We illustrate the usefulness of our results by discussing two examples from a matching calculation and a one-loop basis change.
}

\begin{document}
\maketitle

\clearpage

\section{Introduction}
Loop calculations including four-fermion operators often involve the use of Fierz transformations \cite{Fierz:1939zz}. Applying these identities is however problematic when used in combination with dimensional regularisation, since Fierz identities only hold in $D=4$ space-time dimensions, whereas the loop integral is continued to $D\neq 4$. Consequently, the loop integration and the Fierzing of an operator do not commute. This mismatch between the two operations is usually compensated by introducing evanescent operators \cite{Buras:1989xd,Dugan:1990df,Herrlich:1994kh}. These so-called Fierz-evanescent operators are defined by the difference of an operator and its Fierz-transformed version and therefore vanish in four space-time dimensions, but are non-zero in general $D$ dimensions.

\noindent
In this article we propose a simple way to deal with the issue of Fierz-evanescent contributions, by generalizing the Fierz transformations to the one-loop level. The one-loop corrections to the tree-level relations result from the insertion of Fierz-evanescent operators into one-loop diagrams. These contributions can be computed once and then added as shifts to the tree-level relations. We provide these simple shifts that generalize the standard Fierz relations to include one-loop QCD and QED corrections.
Such one-loop Fierz identities can then be used in the context of one-loop matching computations, in which the resulting operators need to be fierzed in order to obtain the basis of choice. Another application is  the basis change of two-loop anomalous dimension matrices. Such transformations play for instance an important role in the context of the Standard Model Effective Theory (SMEFT) \cite{Grzadkowski:2010es} and its matching onto the Weak Effective Theory (WET) at the electroweak (EW) scale. The SMEFT-WET matching is known at tree-level \cite{Aebischer:2015fzz,Jenkins:2017jig} and since recently also at one-loop \cite{Dekens:2019ept}, and the full one-loop Renormalization Group (RG) running is known above \cite{Alonso:2013hga, Jenkins:2013zja, Jenkins:2013wua} and below \cite{Jenkins:2017dyc,Aebischer:2017gaw} the EW scale. However, in order to cancel the scheme-dependence present in the WET Wilson coefficients, which is introduced via the one-loop matching, the two-loop RG running below the EW scale has to be included. For this reason, in deriving the one-loop Fierz identities we will adopt a generalised BMU scheme, in which Greek identities\footnote{The Greek identities were first used in the context of evanescent operators in \cite{Buras:1989xd}.} \cite{Tracas:1982gp} are used to reduce Dirac structures to the initial basis, but where the scheme-dependent parts (proportional to $\epsilon$) are kept general.\footnote{The original BMU scheme is recovered from the generalized one by setting  $a_{1,2,3}=b_{1,2,3}=c_{1,2,3}=d_{1,2,3}=e_{1,2}=f_{1,2}=1$ in the Greek identities listed in App.  \ref{app:greek}.}$^,$\footnote{We use the $\overline{\text{MS}}$-scheme together with NDR and the convention $D=4-2\epsilon$. Evanescent counterterms in the BMHV scheme have been discussed for instance in \cite{Belusca-Maito:2020ala}.}
The reason for this choice is that most of the known NLO anomalous dimension matrices (ADMs) in the WET are computed in the original BMU scheme: The two-loop QCD corrections for the $\Delta F=1$ and $\Delta F=2$ four-quark operators were computed in \cite{Buras:2000if}. The $\mathcal{O}(\alpha\alpha_s)$ corrections to the SM current-current operators have been derived in \cite{Buras:1992zv,Ciuchini:1993vr,Gilman:1979bc,Shifman:1976ge} and the two-loop ADMs for the SM QCD- and QED penguin operators are given in \cite{Buras:1991jm, Buras:1992tc, Buras:1992zv, Ciuchini:1993vr}.
In order to be able to perform a consistent NLO SMEFT analysis, these results have to be transformed into the JMS basis \cite{Jenkins:2017jig}, in which the one-loop matching is computed. This step has already been performed for four-quark operators in the context of $\Delta F=1$ processes in \cite{Aebischer:2022tvz,Aebischer:2021raf,Aebischer:2021hws} and for $\Delta F=2$ processes in \cite{Aebischer:2022anv}. The generalized Fierz identities derived in this article will however allow to transform the full set of operators at one-loop to the JMS basis. They therefore consist an important step in the pursuit of a full NLO SMEFT analysis.

\noindent
Changing the BMU ADM into the JMS ADM is however a rather special case, since the JMS scheme can easily be transformed into the original BMU one.\footnote{The BMU scheme corresponds to $a_{\text{ev}}=b_{\text{ev}}=c_{\text{ev}}=d_{\text{ev}}=e_{\text{ev}}=f_{\text{ev}}=1$ in \cite{Dekens:2019ept}.} This makes the one-loop basis transformation rather simple, since only Fierz-evanescent operators have to be considered in the basis change. In full generality however the definition of evanescent operators involves general linear combinations of physical operators multiplied by $\epsilon$ \cite{Chetyrkin:1997gb,Gorbahn:2004my}, as opposed to the generalised BMU scheme shown in App.~\ref{app:greek}. In order to be able to change also to bases containing such evanescent operators the renormalization constants $Z_{QQ},\,Z_{QE}$ and $Z_{EE}$ are needed \cite{Gorbahn:2004my}, which we report for Fierz-evanescent operators in App.~\ref{app:Zs}.

\noindent
The rest of the article is organized as follows: in Sec.~\ref{sec:procedure} we illustrate the procedure on how to obtain one-loop corrections of the Fierz identities. We report our results for all possible Dirac structures and fermion field combinations in Sec.~\ref{sec:results}. The usefulness of our results is illustrated in Sec.~\ref{sec:example}, were we show how they can be applied in two different contexts: first in a one-loop matching calculation involving Leptoquarks and secondly in a one-loop basis change of a two-loop ADM governing $\Delta F=2$ processes. Finally we conclude in Sec.~\ref{sec:conclusions}. Several useful results are collected in the Appendices: App.~\ref{app:greek} contains the generalized Greek identities used in this article. In App.~\ref{app:loop} we report all divergent structures encountered in the calculation. App.~\ref{app:genshifts} lists the one-loop corrections to the tree-level Fierz relations in the generalized BMU scheme.
In App.~\ref{app:1loop} we collect one-loop corrections to the physical and Fierz-evanescent operators in our computation and in App.~\ref{app:Zs} we provide the renormalization constants $Z_{QQ},\,Z_{QE}$ and $Z_{EE}$ for bases containing Fierz-evanescent operators.

\section{General Procedure}\label{sec:procedure}

\begin{figure}[H]
 \centering
 \includegraphics[width=8.6cm]{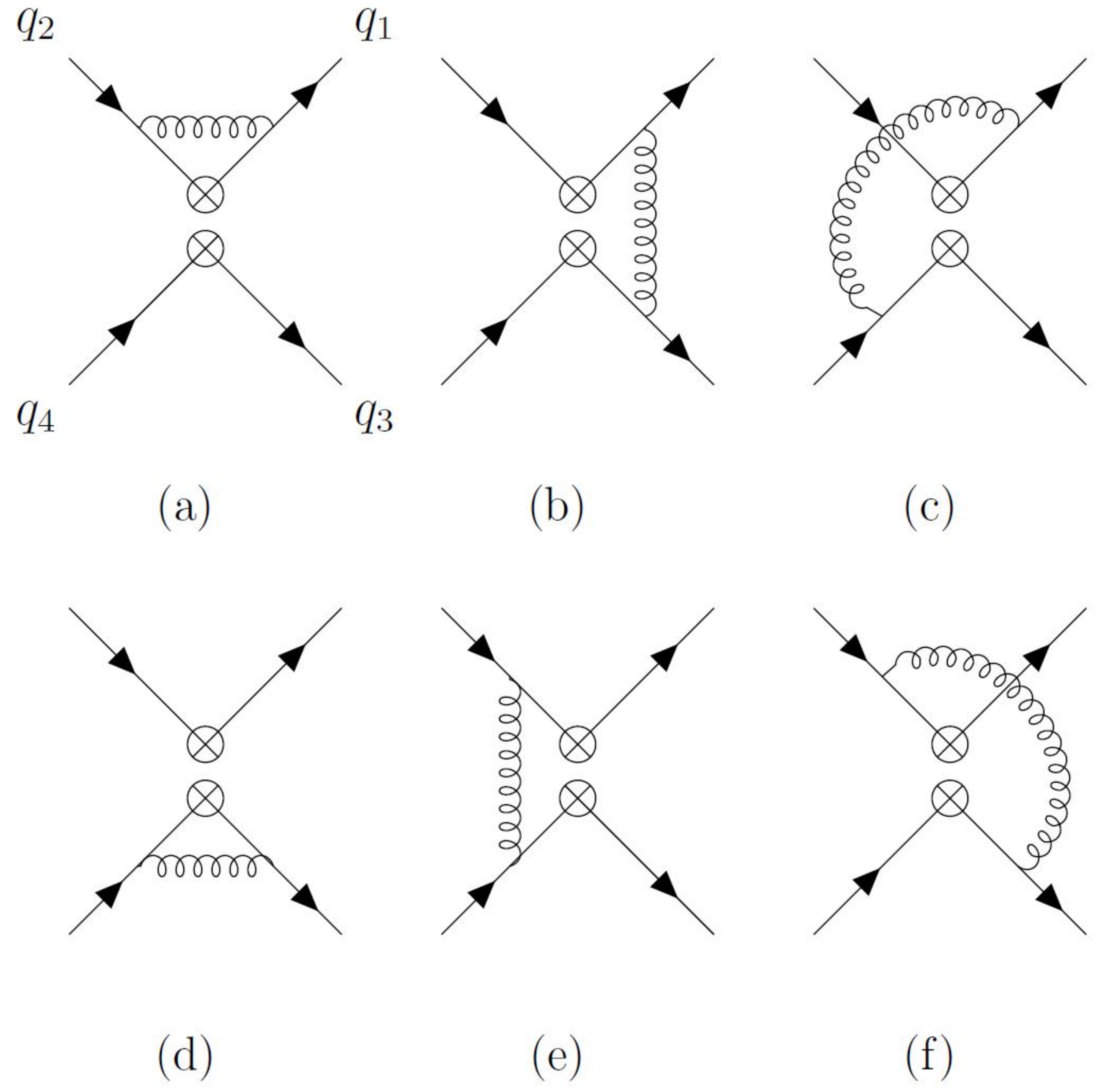}
 \caption{One-loop QCD vertex corrections to four-quark operators. \label{fig:qcdvertex}}
 \end{figure}

\begin{figure}[t]
 \centering
 \includegraphics[width=8.6cm]{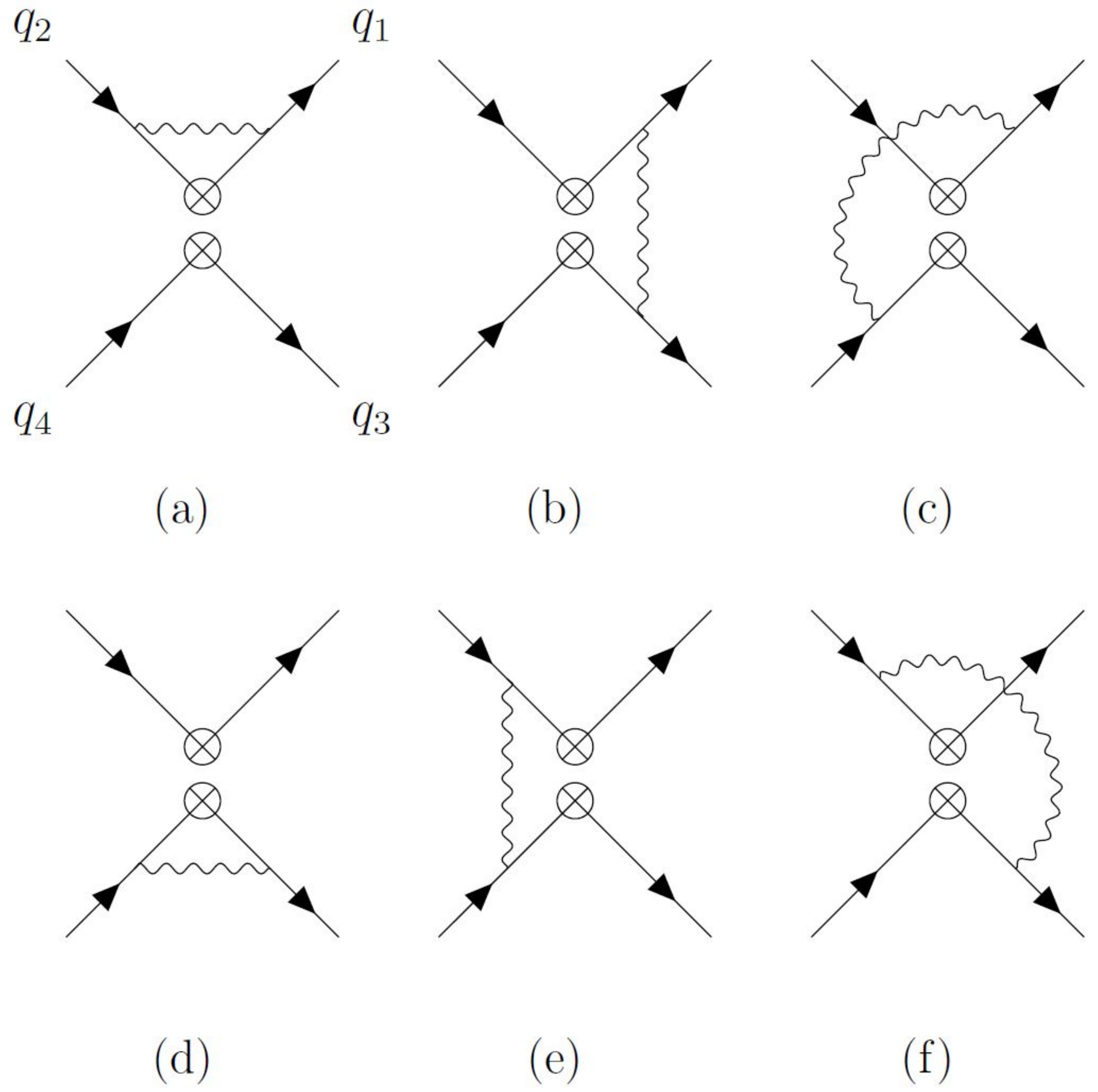}
 \caption{One-loop QED vertex corrections to four-quark operators. \label{fig:qedvertex}}
 \end{figure}

In order to obtain one-loop corrections to Fierz identities let us consider an operator $\mathcal{O}$ and its Fierz-transformed version $\mathcal{FO}$, where $\mathcal{F}$ denotes the application of tree-level Fierz identities. At tree-level the two structures are trivially related:

\begin{equation}
  \mathcal{O} =\mathcal{FO}\,,\qquad (\text{tree-level})
\end{equation}
\noindent
whereas at one-loop the relation has to be generalized by introducing an evanescent operator $E$:

\begin{equation}
  \mathcal{O} =\mathcal{FO}+E\,.\qquad (\text{one-loop})
\end{equation}
The one-loop shift to the tree-level Fierz identity induced by the evanescent operator is then obtained by computing the one-loop corrections ($L$) to the operator and its Fierz-transformed version and taking the difference:
\begin{equation}
  L\mathcal{O} - L\mathcal{FO} = LE\,.
\end{equation}
As an example let us consider a four-quark operator
\begin{equation}
  \mathcal{Q}=(\overline q_1 \Gamma_1 q_2)(\overline q_3 \Gamma_2 q_4)\,,
\end{equation}
with four different quarks $q_{1,2,3,4}$ and two general Dirac structures $\Gamma_{1,2}$. The Fierz-transformed operator has the form
\begin{equation}
  \mathcal{FQ}=(\overline q_1^\alpha \widetilde\Gamma_1 q^\beta_4)(\overline q^\beta_3 \widetilde\Gamma_2 q_2^\alpha)\,,
\end{equation}
with different Dirac structures $\widetilde\Gamma_{1,2}$ and the colour indices $\alpha,\beta$. In order to obtain the shift resulting from the Fierz-evanescent operator
\begin{equation}
  E_{\mathcal{Q}}=\mathcal{Q}-\mathcal{FQ}\,,
\end{equation}
one-loop corrections to the two operators have to be considered. They include genuine vertex corrections as well as contributions from Wavefunction renormalization and operator renormalization. As an example we consider the process $q_2 q_4\to q_1 q_3$, for which the vertex corrections are depicted in Fig.~\ref{fig:qcdvertex} in the case of QCD and in Fig.~\ref{fig:qedvertex} for QED. Since the two operators do not have the same flavour structure, one has to perform an additional (tree-level) Fierz transformation on one of the amplitudes after loop-integration in order to compare the two resulting amplitudes. 
\noindent
This Fierz transformation can be carried out at tree-level, since the amplitude is already at the one-loop order and therefore contributions from higher-order Fierz transformations can be neglected. Also Wavefunction renormalization contributions do not have to be considered, since they drop out in the difference after this additional Fierz transformation.

\noindent
When computing the one-loop corrections to the operators new Dirac structures occur, which have to be reduced to the initial operator basis. To perform this projection we use the generalized Greek identities collected in App~\ref{app:greek}. This choice of how to reduce the new Dirac structures by specifying the $\epsilon$-dependent pieces in the reduction defines the generalized BMU scheme, for which we report the obtained shifts in App.~\ref{app:genshifts}. The original BMU scheme has been introduced in \cite{Buras:2000if} and its Greek identities can be found in \cite{Buras:2012fs}. Since the shifts from evanescent operators precisely result from these $\epsilon$-dependent parts of the projections, only the poles of the one-loop diagrams have to be computed. The insertion of an evanescent operator into these divergent structures will then yield the finite shift in the Fierz identities.

\begin{figure}[tbp]
     \centering
     \begin{subfigure}{0.59\textwidth}
         \centering
         \includegraphics[width=2.9cm]{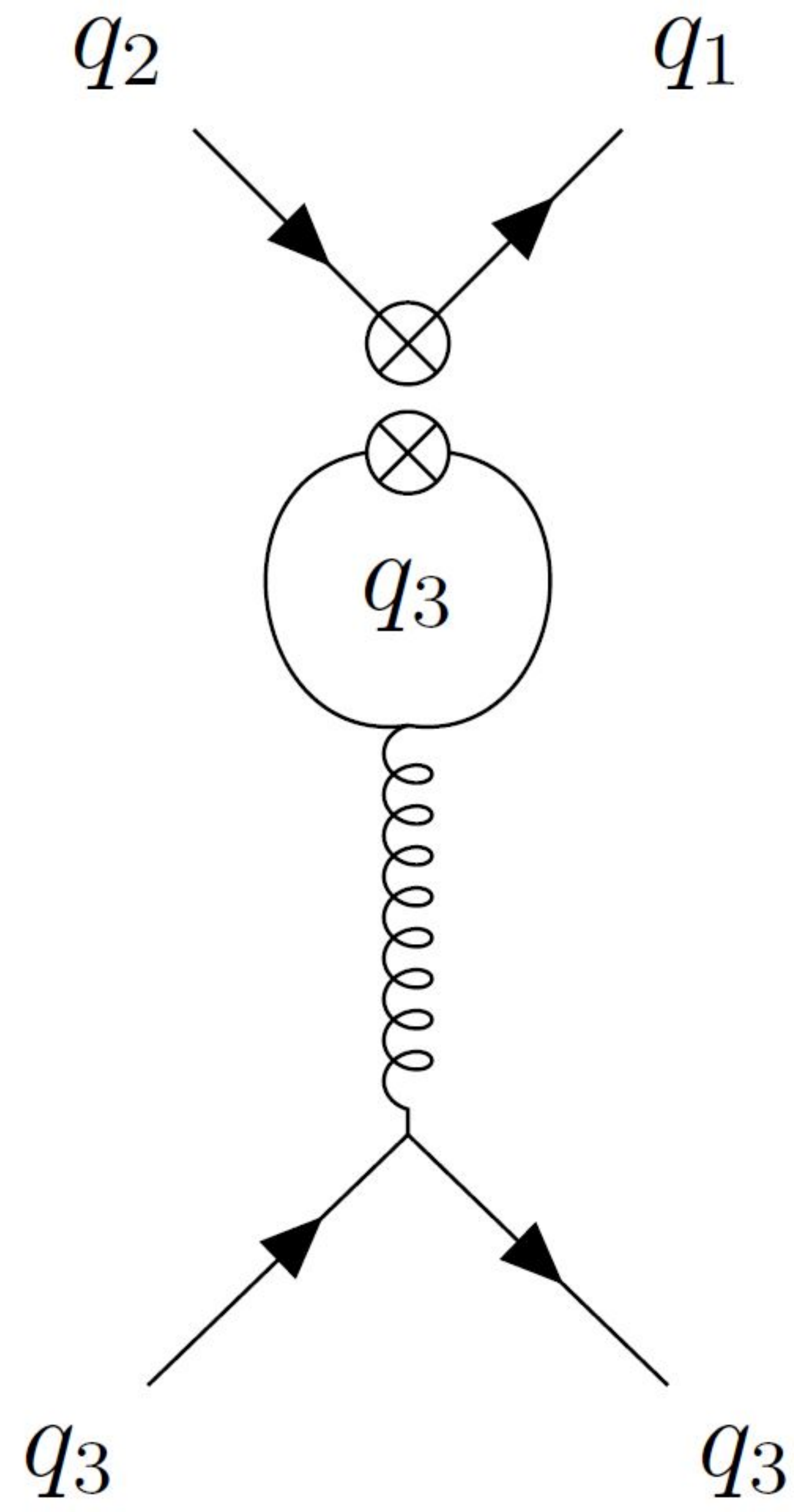}
         \caption{}
         \label{fig:closedP}
     \end{subfigure}
    \hspace*{-4cm}
     \begin{subfigure}{0.59\textwidth}
		\centering
		\vspace*{+0.3cm}
         \includegraphics[width=2.9cm]{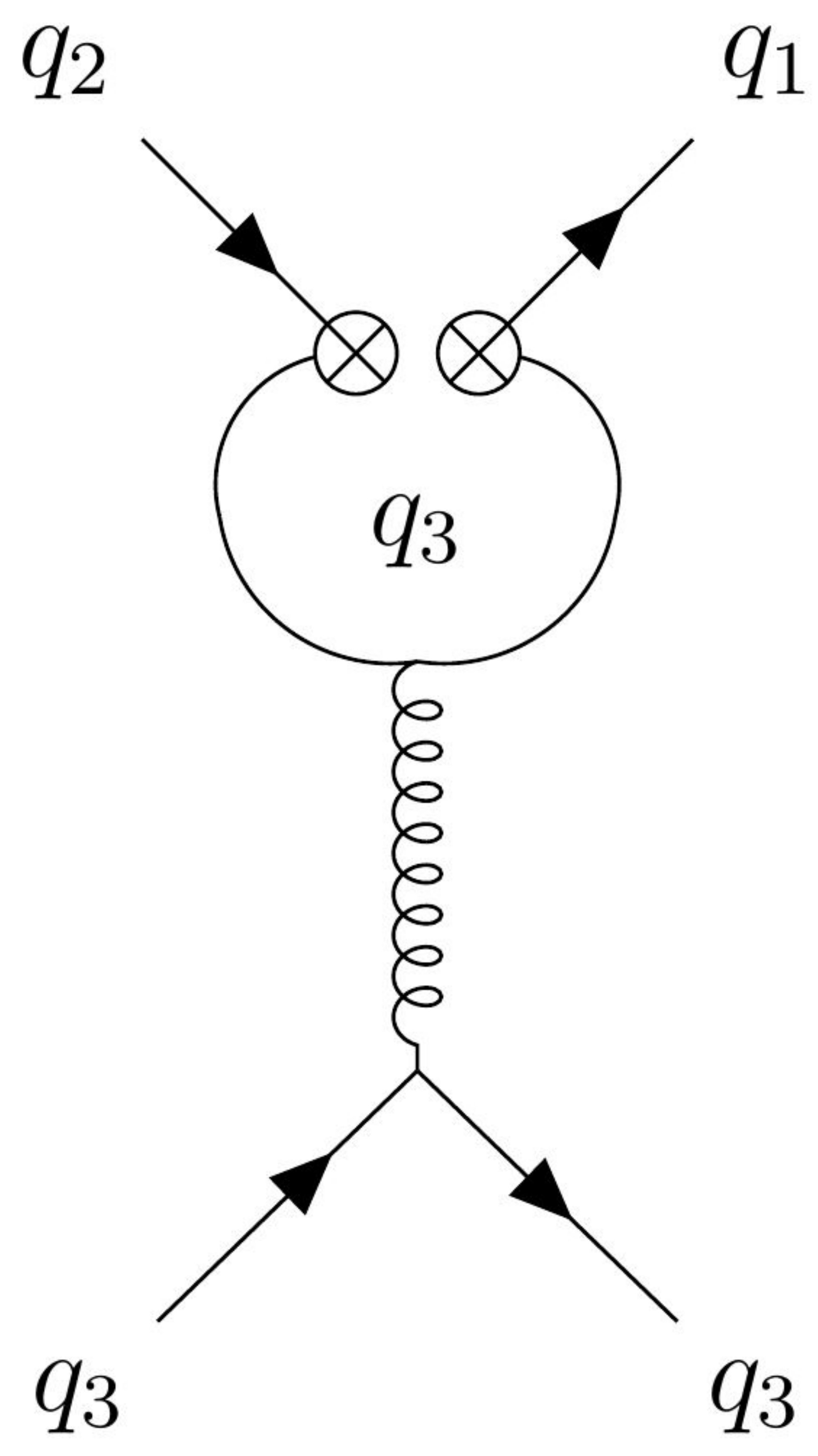}
         \caption{}
         \label{fig:openP}
     \end{subfigure}
        \caption{QCD penguin diagrams for four-quark operators.
        }
        \label{fig:qcdpeng}
\end{figure}

\noindent
If two or more fermions of a four-fermi operator are the same, then also penguin diagrams have to be taken into account. Consider a four-quark operator of the form:
\begin{equation}
  \mathcal{P} = (\overline q_1 \Gamma_1 q_2) (\overline q_3 \Gamma_2 q_3)\,,
\end{equation}
and its Fierz-transformed version
\begin{equation}
  \mathcal{FP} = (\overline q^\alpha_1 \widetilde\Gamma_1 q^\beta_3) (\overline q^\beta_3 \widetilde\Gamma_2 q^\alpha_2)\,,
\end{equation}
then $\mathcal{P}$ generates closed penguin diagrams and $\mathcal{FP}$ open penguin diagrams, which for QCD are shown in Fig.~\ref{fig:qcdpeng}. Similar diagrams are generated when QED corrections are considered.

\noindent
In Appendix~\ref{app:loop} we report all divergent structures that can be obtained from vertex corrections and penguin diagrams. These structures, together with the Greek identities will then be used in the next section to obtain one-loop corrections to Fierz relations.

\section{Results}\label{sec:results}
In this section we report the results obtained, following the procedure outlined in the previous section. We compute the one-loop QCD- and QED-shifts to the tree-level Fierz relations for general four-fermi operators with vector-, scalar- and tensor Dirac structures and different combinations of fermion fields.
All results obtained in this section are also valid for the Parity transformed operators which result from the interchange $P_L\leftrightarrow P_R$, since QCD and QED are invariant under Parity transformations. In order to simplify the presentation we will adopt the original BMU scheme in the following. The general results computed in the generalized BMU scheme are collected in App.~\ref{app:genshifts}.


\subsection{Vector Operators}

In this subsection we discuss the shifts for four-fermi vector operators, for which we use the following notation\footnote{In addition we use the shorthand notation $  \op{V}{\1B}{f_if_jf_kf_l}\equiv(\overline f_i \gamma_\mu f_j)(\overline f_k \gamma^\mu P_B f_l)\,,$ and analogous expressions.}:

\begin{equation}
  \op{V}{AB}{f_if_jf_kf_l}\equiv(\overline f_i \gamma_\mu P_A f_j)(\overline f_k \gamma^\mu P_B f_l)\,,  \quad \text{ where } f=q,\ell \,.
\end{equation}
For four-quark operators we define in addition the colour-crossed operators:
\begin{equation}
  \cop{V}{AB}{q_iq_jq_kq_l}\equiv(\overline q_i^\alpha \gamma_\mu P_A q_j^\beta)(\overline q_k^\beta \gamma^\mu P_B q_l^\alpha)  \,,
\end{equation}
where $\alpha,\beta$ denote colour indices.

\noindent
The resulting shifts for vector operators are collected for four-quark operators in Tab.~\ref{tab:V4quark}, for semi-leptonic operators in Tab.~\ref{tab:VSL} and for four-lepton operators in Tab.~\ref{tab:V4lepton}. Since the genuine vertex corrections of a four-fermi vector operator are the same as for its Fierz-transformed version, the first lines in the above mentioned tables vanish. Consequently, only operators with at least two equal fermions obtain a shift at one-loop, which results from penguin contributions. Finally, four-lepton operators with four equal leptons do not obtain any one-loop shift, since their Fierz-transformed version is equal to the initial operator. Our results for the QCD shift of $\Delta F=1$ four-quark operators agree with the findings in \cite{Aebischer:2022tvz} and for $\Delta F=2$ with \cite{Aebischer:2020dsw}.

\begin{table}[tbp]
\renewcommand{\arraystretch}{1.5}
\small
\begin{align*}
\begin{array}[t]{c|c|c|c}
\toprule
 \text{Operator} & \text{Tree-level Fierz} & \text{QCD shift} &  \text{QED shift}   \\
\midrule\midrule
\op{V}{LL}{q_1q_2q_3q_4} & \cop{V}{LL}{q_1q_4q_3q_2} & 0 & 0   \\
\op{V}{LL}{q_1q_2q_3q_3} & \cop{V}{LL}{q_1q_3q_3q_2} & 0 & -\frac{2N_c}{3}Q_{q_3}^2 \op{V}{L\1}{q_1q_2q_3q_3}  \\
\cop{V}{LL}{q_1q_2q_3q_3} & \op{V}{LL}{q_1q_3q_3q_2} & -\frac{1}{3} \cop{V}{L\1}{q_1q_2q_3q_3} + \frac{1}{3N_c}\op{V}{L\1}{q_1q_2q_3q_3} & -\frac{2}{3}Q_{q_3}^2 \op{V}{L\1}{q_1q_2q_3q_3}   \\
\op{V}{LL}{q_1q_1q_2q_2} & \cop{V}{LL}{q_1q_2q_2q_1} & 0 & -\frac{2N_c}{3} \left(Q_{q_1}^2\op{V}{\1L}{q_1q_1q_2q_2}+Q_{q_2}^2\op{V}{L\1}{q_1q_1q_2q_2}\right)   \\
\cop{V}{LL}{q_1q_1q_2q_2} & \op{V}{LL}{q_1q_2q_2q_1} & -\frac{1}{3} \cop{V}{L\1}{q_1q_1q_2q_2} + \frac{1}{3N_c}\op{V}{L\1}{q_1q_1q_2q_2}+L\leftrightarrow \1 & -\frac{2}{3} \left(Q_{q_1}^2\op{V}{\1L}{q_1q_1q_2q_2}+Q_{q_2}^2\op{V}{L\1}{q_1q_1q_2q_2}\right)   \\
\op{V}{LL}{q_1q_1q_2q_1} & \cop{V}{LL}{q_1q_1q_2q_1} & \frac{1}{3} \cop{V}{\1L}{q_1q_1q_2q_1} - \frac{1}{3N_c}\op{V}{\1L}{q_1q_1q_2q_1} &  - \frac{2}{3}(N_c-1)Q_{q_1}^2 \op{V}{\1L}{q_1q_1q_2q_1}   \\
\op{V}{LL}{q_1q_1q_1q_1} & \cop{V}{LL}{q_1q_1q_1q_1} & \frac{2}{3} \cop{V}{L\1}{q_1q_1q_1q_1} - \frac{2}{3N_c}\op{V}{L\1}{q_1q_1q_1q_1} &  - \frac{4}{3}(N_c-1)Q_{q_1}^2 \op{V}{L\1}{q_1q_1q_1q_1}  \\
\bottomrule
\end{array}
\end{align*}
\caption{Fierz transformations for four-quark $\gamma_\mu P_L \otimes \gamma^\mu P_L$ operators together with their one-loop QCD and QED shifts in the original BMU scheme.}
\label{tab:V4quark}
\end{table}

\begin{table}[tbp]
\renewcommand{\arraystretch}{1.5}
\small
\begin{align*}
\begin{array}[t]{c|c|c|c}
\toprule
 \text{Operator} & \text{Tree-level Fierz} & \text{QCD shift} &  \text{QED shift}   \\
\midrule\midrule
\op{V}{LL}{q_1q_2\ell_1\ell_2} & \op{V}{LL}{q_1\ell_2\ell_1q_2} & 0 & 0 \\
\op{V}{LL}{q_1q_1\ell_1\ell_2} & \op{V}{LL}{q_1\ell_2\ell_1q_1} & 0 & -\frac{2N_c}{3}Q_{q_1}^2 \op{V}{\1L}{q_1q_1\ell_1\ell_2} \\
\op{V}{LL}{q_1q_2\ell_1\ell_1} & \op{V}{LL}{q_1\ell_1\ell_1q_2} & 0 & -\frac{2}{3}Q_{\ell_1}^2 \op{V}{L\1}{q_1q_2\ell_1\ell_1} \\
\op{V}{LL}{q_1q_1\ell_1\ell_1} & \op{V}{LL}{q_1\ell_1\ell_1q_1} & 0 & -\frac{2}{3}\left( Q_{\ell_1}^2 \op{V}{L\1}{q_1q_1\ell_1\ell_1}+N_cQ_{q_1}^2 \op{V}{\1L}{q_1q_1\ell_1\ell_1}\right) \\
\bottomrule
\end{array}
\end{align*}
\caption{Fierz transformations for semi-leptonic $\gamma_\mu P_L \otimes \gamma^\mu P_L$ operators together with their one-loop QCD and QED shifts in the original BMU scheme.}
\label{tab:VSL}
\end{table}

\begin{table}[tbp]
\renewcommand{\arraystretch}{1.5}
\small
\begin{align*}
\begin{array}[t]{c|c|c|c}
\toprule
 \text{Operator} & \text{Tree-level Fierz} & \text{QCD shift} &  \text{QED shift}   \\
\midrule\midrule
\op{V}{LL}{\ell_1\ell_2\ell_3\ell_4} & \cop{V}{LL}{\ell_1\ell_4\ell_3\ell_2} & 0 & 0   \\
\op{V}{LL}{\ell_1\ell_1\ell_2\ell_3} & \op{V}{LL}{\ell_1\ell_3\ell_2\ell_1} & 0 & -\frac{2}{3}Q_{\ell_1}^2 \op{V}{\1L}{\ell_1\ell_1\ell_2\ell_3}\\
\op{V}{LL}{\ell_1\ell_1\ell_2\ell_2} & \op{V}{LL}{\ell_1\ell_2\ell_2\ell_1} & 0 & -\frac{2}{3}Q_{\ell_1}^2 \op{V}{\1L}{\ell_1\ell_1\ell_2\ell_2}-\frac{2}{3}Q_{\ell_2}^2 \op{V}{L\1}{\ell_1\ell_1\ell_2\ell_2}\\
\op{V}{LL}{\ell_1\ell_1\ell_1\ell_2} & \op{V}{LL}{\ell_1\ell_2\ell_1\ell_1} & 0 & 0\\
\op{V}{LL}{\ell_1\ell_1\ell_1\ell_1} & \op{V}{LL}{\ell_1\ell_1\ell_1\ell_1} & 0 & 0\\
\bottomrule
\end{array}
\end{align*}
\caption{Fierz transformations for four-lepton $\gamma_\mu P_L \otimes \gamma^\mu P_L$ operators together with their one-loop QCD and QED shifts in the original BMU scheme.}
\label{tab:V4lepton}
\end{table}

\newpage
\subsection{Scalar and Vector LR Operators}
In this subsection we discuss scalar and vector four-fermi operators with the Dirac structures $P_L \otimes P_R$ and $\gamma_\mu P_R \otimes \gamma^\mu P_L$, respectively. At tree-level these types of operators are related through a Fierz relation. In addition to the vector operators defined in the previous subsection we define the analogous scalar operators as
\begin{equation}
\op{S}{AB}{f_1f_2f_3f_4} \equiv (\overline f_1 P_A f_2)(\overline f_3 P_B f_4)\, .
\end{equation}
These operators do not obtain any shift at the one-loop level, neither from QCD nor from QED, which is indicated in Tab.~\ref{tab:SLR}. The genuine vertex corrections drop out in the difference, which was also the case for VLL\footnote{We use the shorthand notation VLL to denote vector operators with $\gamma_\mu P_L \otimes \gamma^\mu P_L$ Dirac structure and analogous abbreviations for scalar and tensor operators.} operators in the previous subsection. In addition, also the divergent parts of the penguin diagrams either vanish or cancel in the difference. Consequently, all shifts vanish for the LR operators, independently of the fermion combinations.
This is a special feature of the original BMU scheme and does not hold anymore in the generalized BMU scheme, as can be seen in Tab.~\ref{tab:genBMULR} in the Appendix.
\begin{table}[tbp]
\renewcommand{\arraystretch}{1.5}
\small
\begin{align*}
\begin{array}[t]{c|c|c|c}
\toprule
 \text{Operator} & \text{Tree-level Fierz} & \text{QCD shift} &  \text{QED shift}   \\
\midrule\midrule
\op{S}{LR}{f_1f_2f_3f_4} &-\frac{1}{2} \cop{V}{RL}{f_1f_4f_3f_2}&0& 0\\
\cop{S}{LR}{f_1f_2f_3f_4} &-\frac{1}{2} \op{V}{RL}{f_1f_4f_3f_2}&0& 0\\
\bottomrule
\end{array}
\end{align*}
\caption{Fierz transformations for scalar $P_L \otimes P_R$ operators together with their one-loop QCD and QED shifts in the original BMU scheme.}
\label{tab:SLR}
\end{table}

\subsection{Scalar and Tensor LL Operators}
Finally we discuss the scalar and tensor operators with LL structure. Besides the notation for scalar operators defined in the previous subsection we define:
\begin{equation}
\op{T}{LL}{f_1f_2f_3f_4} \equiv (\overline f_1 \sigma_{\mu \nu} P_L f_2)(\overline f_3 \sigma^{\mu \nu} P_L f_4)\,, \quad \text{ where } \sigma^{\mu \nu} \equiv \frac{i}{2}\left[ \gamma^\mu, \gamma^\nu \right] \, .
\end{equation}
and for colour-crossed four-quark operators we use:
\begin{equation}
\cop{T}{LL}{q_1q_2q_3q_4} \equiv (\overline q_1^\alpha \sigma_{\mu \nu} P_L q_2^\beta)(\overline q_3^\beta \sigma^{\mu \nu} P_L q_4^\alpha)\,.
\end{equation}
For these operators the shifts result exclusively from genuine vertex corrections, since the penguin diagrams again vanish, which can be deduced from the results in App.~\ref{app:loop}. For that reason the QED corrections have the same form for all fermion operators, since they only depend on the fermion charges. To shorten the notation we define:

\begin{equation}
Q_{ijkl}\equiv Q_i Q_j + Q_k Q_l \, .
\end{equation}
Furthermore the first and fourth diagrams in Fig.~\ref{fig:qcdvertex} and Fig.~\ref{fig:qedvertex} are finite for tensor operators and therefore do not have to be considered.

\noindent
We report the one-loop shifts for scalar operators in Tab.~\ref{tab:scalar} and for tensor operators in Tab.~\ref{tab:tensor}. Our QCD results for $\Delta F=1$ and $\Delta F=2$ operators again agree with the ones obtained in \cite{Aebischer:2022tvz,Aebischer:2020dsw}. Several entries in Tabs.~\ref{tab:scalar} and \ref{tab:tensor} are redundant, since the scalar and tensor operators mix under Fierz transformations. For convenience we report however all entries for scalar and tensor operators and remind the reader of this redundancy.

\begin{table}[tbp]
\renewcommand{\arraystretch}{1.9}
\small
\begin{adjustwidth}{-1.1cm}{}
\begin{align*}
\begin{array}[t]{c|c|c|c}
\toprule
 \text{Operator} & \text{Tree-level Fierz} & \text{QCD shift} &  \text{QED shift}   \\
\midrule\midrule
\op{S}{LL}{q_1q_2q_3q_4}
&-\frac{1}{2}\cop{S}{LL}{q_1q_4q_3q_2}  - \frac{1}{8}\cop{T}{LL}{q_1q_4q_3q_2}
& \makecell{-\frac{1}{N_c}\op{S}{LL}{q_1q_2q_3q_4}+\cop{S}{LL}{q_1q_2q_3q_4}\\+\frac{N_c^2-6}{8N_c}\op{T}{LL}{q_1q_2q_3q_4}+\frac{5}{8}\cop{T}{LL}{q_1q_2q_3q_4}}
& \makecell{\frac{1}{2}(Q_1 +Q_2)(Q_3+Q_4)\op{S}{LL}{q_1q_2q_3q_4} \\ +\frac{1}{8}\left(Q_{1234}+2Q_{1423}+3Q_{1324}\right)\op{T}{LL}{q_1q_2q_3q_4}\\ \nonumber} \\
\cop{S}{LL}{q_1q_2q_3q_4}
&-\frac{1}{2}\op{S}{LL}{q_1q_4q_3q_2}  - \frac{1}{8}\op{T}{LL}{q_1q_4q_3q_2}
& \makecell{\frac{1}{2}\op{S}{LL}{q_1q_2q_3q_4}+\frac{N_c^2-2}{2N_c}\cop{S}{LL}{q_1q_2q_3q_4}\\+\frac{1}{2}\op{T}{LL}{q_1q_2q_3q_4}+\frac{N_c^2-3}{4N_c}\cop{T}{LL}{q_1q_2q_3q_4}}
& \makecell{\frac{1}{2}(Q_1 +Q_2)(Q_3+Q_4)\cop{S}{LL}{q_1q_2q_3q_4} \\ +\frac{1}{8}\left(Q_{1234}+2Q_{1423}+3Q_{1324}\right)\cop{T}{LL}{q_1q_2q_3q_4}\\ \nonumber} \\
\op{S}{LL}{q_1q_2\ell_1\ell_2}
&-\frac{1}{2}\op{S}{LL}{q_1\ell_2\ell_1q_2}  - \frac{1}{8}\op{T}{LL}{q_1\ell_2\ell_1q_2}
& \frac{N_c^2-1}{16N_c}\op{T}{LL}{q_1q_2\ell_1\ell_2}
& \makecell{\frac{1}{2}(Q_1 +Q_2)(Q_3+Q_4)\op{S}{LL}{q_1q_2\ell_1\ell_2} \\ +\frac{1}{8}\left(Q_{1234}+2Q_{1423}+3Q_{1324}\right)\op{T}{LL}{q_1q_2\ell_1\ell_2}\\ \nonumber} \\
\op{S}{LL}{q_1\ell_2\ell_1q_2}
&-\frac{1}{2}\op{S}{LL}{q_1q_2\ell_1\ell_2}  - \frac{1}{8}\op{T}{LL}{q_1q_2\ell_1\ell_2}
& \frac{7-7N_c^2}{8N_c}\op{S}{LL}{q_1\ell_2\ell_1q_2}+\frac{N_c^2-1}{32N_c}\op{T}{LL}{q_1\ell_2\ell_1q_2}
& \makecell{\frac{1}{2}(Q_1 +Q_2)(Q_3+Q_4)\op{S}{LL}{q_1\ell_2\ell_1q_2} \\ +\frac{1}{8}\left(Q_{1234}+2Q_{1423}+3Q_{1324}\right)\op{T}{LL}{q_1\ell_2\ell_1q_2}\\ \nonumber} \\
\op{S}{LL}{\ell_1\ell_2\ell_3\ell_4}
&-\frac{1}{2}\op{S}{LL}{\ell_1\ell_4\ell_3\ell_2}  - \frac{1}{8}\op{T}{LL}{\ell_1\ell_4\ell_3\ell_2}
& 0
& \makecell{\frac{1}{2}(Q_1 +Q_2)(Q_3+Q_4)\op{S}{LL}{\ell_1\ell_2\ell_3\ell_4} \\ +\frac{1}{8}\left(Q_{1234}+2Q_{1423}+3Q_{1324}\right)\op{T}{LL}{\ell_1\ell_2\ell_3\ell_4}\\ \nonumber} \\
\bottomrule
\end{array}
\end{align*}
\end{adjustwidth}
\caption{Fierz transformations for scalar $P_L \otimes P_L$ operators together with their one-loop QCD and QED shifts in the original BMU scheme.}
\label{tab:scalar}
\end{table}

\begin{table}[tbp]
\renewcommand{\arraystretch}{1.9}
\small
\begin{adjustwidth}{-1.2cm}{}
\begin{align*}
\begin{array}[t]{c|c|c|c}
\toprule
 \text{Operator} & \text{Tree-level Fierz} & \text{QCD shift} &  \text{QED shift}   \\
\midrule\midrule
\op{T}{LL}{q_1q_2q_3q_4} &
-6\cop{S}{LL}{q_1q_4q_3q_2}  + \frac{1}{2}\cop{T}{LL}{q_1q_4q_3q_2}  &
\makecell{\frac{44-14N_c^2}{N_c}\op{S}{LL}{q_1q_2q_3q_4}-30\cop{S}{LL}{q_1q_2q_3q_4}\\+\frac{1}{N_c}\op{T}{LL}{q_1q_2q_3q_4}-\cop{T}{LL}{q_1q_2q_3q_4}} &
\makecell{-2\left(4Q_{1423}+7Q_{1234}+11Q_{1324}\right)\op{S}{LL}{q_1q_2q_3q_4} \\ -\frac{1}{2}(Q_1 +Q_2)(Q_3+Q_4)\op{T}{LL}{q_1q_2q_3q_4}\\ \nonumber} \\
\cop{T}{LL}{q_1q_2q_3q_4} &
-6\op{S}{LL}{q_1q_4q_3q_2}  + \frac{1}{2}\op{T}{LL}{q_1q_4q_3q_2}  &
\makecell{-36\op{S}{LL}{q_1q_2q_3q_4}+\frac{44-8N_c^2}{N_c}\cop{S}{LL}{q_1q_2q_3q_4}\\-\frac{1}{2}\op{T}{LL}{q_1q_2q_3q_4}+\frac{2-N_c^2}{2N_c}\cop{T}{LL}{q_1q_2q_3q_4}} &
\makecell{-2\left(4Q_{1423}+7Q_{1234}+11Q_{1324}\right)\cop{S}{LL}{q_1q_2q_3q_4} \\ -\frac{1}{2}(Q_1 +Q_2)(Q_3+Q_4)\cop{T}{LL}{q_1q_2q_3q_4}\\ \nonumber} \\
\op{T}{LL}{q_1q_2\ell_1\ell_2} &
-6\op{S}{LL}{q_1\ell_2\ell_1q_2}  + \frac{1}{2}\op{T}{LL}{q_1\ell_2\ell_1q_2}  &
\frac{7-7N_c^2}{N_c}\op{S}{LL}{q_1q_2\ell_1\ell_2} &
\makecell{-2\left(4Q_{1423}+7Q_{1234}+11Q_{1324}\right)\op{S}{LL}{q_1q_2\ell_1\ell_2} \\ -\frac{1}{2}(Q_1 +Q_2)(Q_3+Q_4)\op{T}{LL}{q_1q_2\ell_1\ell_2}\\ \nonumber} \\
\op{T}{LL}{q_1\ell_2\ell_1q_2} &
-6\op{S}{LL}{q_1q_2\ell_1\ell_2}  + \frac{1}{2}\op{T}{LL}{q_1q_2\ell_1\ell_2}  &
\frac{7N_c^2-7}{2N_c}\op{S}{LL}{q_1\ell_2\ell_1q_2} +\frac{3N_c^2-3}{8N_c}\op{T}{LL}{q_1\ell_2\ell_1q_2} &
\makecell{-2\left(4Q_{1423}+7Q_{1234}+11Q_{1324}\right)\op{S}{LL}{q_1\ell_2\ell_1q_2} \\ -\frac{1}{2}(Q_1 +Q_2)(Q_3+Q_4)\op{T}{LL}{q_1\ell_2\ell_1q_2}\\ \nonumber} \\
\op{T}{LL}{\ell_1\ell_2\ell_3\ell_4} &
-6\op{S}{LL}{\ell_1\ell_4\ell_3\ell_2}  + \frac{1}{2}\op{T}{LL}{\ell_1\ell_4\ell_3\ell_2}  &
0 &
\makecell{-2\left(4Q_{1423}+7Q_{1234}+11Q_{1324}\right)\op{S}{LL}{\ell_1\ell_2\ell_3\ell_4} \\ -\frac{1}{2}(Q_1 +Q_2)(Q_3+Q_4)\op{T}{LL}{\ell_1\ell_2\ell_3\ell_4}\\ \nonumber} \\
\bottomrule
\end{array}
\end{align*}
\end{adjustwidth}
\caption{Fierz transformations for tensor $\sigma_{\mu \nu} P_L \otimes \sigma^{\mu \nu} P_L$ operators together with their one-loop QCD and QED shifts in the original BMU scheme.}
\label{tab:tensor}
\end{table}

\section{Application}\label{sec:example}

In this section we demonstrated the usefulness of the obtained shifts and also how to apply them by studying two examples from the existing literature. First we consider a one-loop matching calculation, in which a Fierz transformation is needed. Secondly we discuss a one-loop basis change of a two-loop anomalous dimension matrix for $\Delta F=2$ transitions, in which the operators differ by a Fierz transformation.

\subsection{Matching}
In this subsection we apply our method to the matching results obtained in \cite{Aebischer:2018acj}, where a general scalar Leptoquark (LQ) was integrated out at the one-loop level in QCD. The resulting semi-leptonic operators are obtained in the LQ basis, meaning that each current of the four-fermi operators contains a quark and a lepton. As an example we consider the scalar and tensor operators in the LQ basis, which in the notation of \cite{Aebischer:2018acj} read

\begin{align}
  \widetilde O_S^{AB} &= (\overline q P_A \ell) (\overline \ell P_B q)\,,\\
  \widetilde O_T^{A} &= (\overline q\sigma_{\mu\nu} P_A \ell) (\overline \ell\sigma^{\mu\nu} P_A q)\,.
\end{align}
For most purposes it is however more favorable to work in the SM basis, which has the following form:
\begin{align}
   O_S^{AB} &= (\overline q P_A q) (\overline \ell P_B \ell)\,,\\
   O_T^{A} &= (\overline q\sigma_{\mu\nu} P_A q) (\overline \ell\sigma^{\mu\nu} P_A \ell)\,.
\end{align}
At the operator level using the QCD relations for the semi-leptonic operators in Tabs.~\ref{tab:scalar} and \ref{tab:tensor} one finds the one-loop relation between the two bases:

\begin{equation}\label{eq:matchrel}
  \left(\begin{array}{c}
              O_S^{AA} \\
           O_T^{A} \\
         \end{array} \right)= R_{0}\left(\begin{array}{c}
               \widetilde O_S^{AA} \\
               \widetilde  O_T^{A} \\
             \end{array} \right)+\frac{\alpha_s}{4\pi}R_{1}\left(\begin{array}{c}
                     O_S^{AA} \\
                  O_T^{A} \\
                \end{array} \right)
\end{equation}
with the matrices
\begin{equation}
  R_0=\left(
  \begin{array}{ccc}
    -\frac{1}{2} & -\frac{1}{8} \\
    -6 & \frac{1}{2} \\
  \end{array}
\right)\,,\qquad R_1= \left(\begin{array}{cc}
    0 & \frac{N_c^2-1}{16N_c} \\
    \frac{7-7N_c^2}{N_c} & 0 \\
  \end{array}\right)\,.
\end{equation}
Solving the relation in eq.~\eqref{eq:matchrel} for the SM basis and expanding in $\alpha_s$ one finds:

\begin{equation}\label{eq:SM_LQ}
  \left(\begin{array}{c}
              O_S^{AA} \\
           O_T^{A} \\
         \end{array} \right)= \left[R_{0}+\frac{\alpha_s}{4\pi}R_1R_{0}\right]\left(\begin{array}{c}
               \widetilde O_S^{AA} \\
               \widetilde  O_T^{A} \\
             \end{array} \right)\,.
\end{equation}
To obtain the corresponding transformation for Wilson coefficients one simply has to take the inverse-transpose of the transformation in eq.~\eqref{eq:SM_LQ}. Expanding in $\alpha_s$ one finds:

\begin{align}\label{eq:WCtrafo}
  \left(\begin{array}{c}
              C_S^{AA} \\
           C_T^{A} \\
         \end{array} \right)&= \left[R_{0}+\frac{\alpha_s}{4\pi}R_{1}R_0\right]^{-T}\left(\begin{array}{c}
                      \widetilde C_S^{AA} \\
                      \widetilde  C_T^{A} \\
                    \end{array} \right) \notag\\
                    &= \left[R_{0}^{-T}-\frac{\alpha_s}{4\pi}R_{1}^TR_{0}^{-T}\right]\left(\begin{array}{c}
                                 \widetilde C_S^{AA} \\
                                 \widetilde  C_T^{A} \\
                               \end{array} \right) \notag\\
  &=\left[\left(
  \begin{array}{ccc}
    -\frac{1}{2} & -6 \\
    -\frac{1}{8} & \frac{1}{2} \\
  \end{array}
\right)
+\frac{\alpha_s}{4\pi}
\left(\begin{array}{cc}
  -\frac{7}{4}C_F & 0 \\
  \frac{1}{16}C_F & 0 \\
\end{array}\right)
\right]
\left(\begin{array}{c}
           \widetilde C_S^{AA} \\
           \widetilde  C_T^{A} \\
         \end{array} \right)\,,
\end{align}
where we used $C_F=\frac{N_c^2-1}{2N_c}$ and set the second column of the second matrix to zero, since the Wilson coefficient of the tensor operator starts at $\mathcal{O}(\alpha_s)$, as can be seen in eq.~(24) of \cite{Aebischer:2018acj}. The transformation in eq.~\eqref{eq:WCtrafo} corresponds exactly to the results in eq.~(26) of \cite{Aebischer:2018acj} and shows the advantage of our method. Having the one-loop Fierz relations at hand allows to perform the matching onto the LQ basis and fierzing them at one-loop, without the need to compute the contributions from Fierz-evanescent operators. The one-loop basis change is therefore reduced to a simple algebraic problem. Note however, that the BMU scheme was used in \cite{Aebischer:2018acj}, which makes the transformation particularly simple. We will discuss a more general example in the next subsection.

\subsection{Basis change}
In appendix C.1 of \cite{Gorbahn:2009pp} a one-loop basis change was performed for $\Delta F=2$ operators an subsequently the two-loop QCD ADM was derived in the new basis. The initial basis in which the two-loop ADM was already known \cite{Buras:2000if} is given by:

\begin{align}
  \overline Q_1^{\text{SLL}} &= (\overline b_R q_L)(\overline b_Rq_L)\,, \\
  \overline Q_2^{\text{SLL}} &= -(\overline b_R\sigma_{\mu\nu} q_L)(\overline b_R\sigma^{\mu\nu} q_L)\,, \\
  \overline E_1^{\text{SLL}} &= (\overline b^i_R q^j_L)(\overline b^j_Rq^i_L)+\frac{1}{2}\overline Q_1^{\text{SLL}}-\frac{1}{8}\overline Q_2^{\text{SLL}}\,, \\
  \overline E_2^{\text{SLL}} &= -(\overline b^i_R\sigma_{\mu\nu} q^j_L)(\overline b^j_R\sigma^{\mu\nu} q^i_L)-6\overline Q_1^{\text{SLL}}-\frac{1}{2}\overline Q_2^{\text{SLL}}\,,
\end{align}
where $i,j$ denote colour-indices. This basis is then translated into the operator basis
\begin{align}
  Q_1^{\text{SLL}} &= (\overline b_R q_L)(\overline b_Rq_L)\,, \\
  \widetilde Q_1^{\text{SLL}} &= (\overline b^i_R q^j_L)(\overline b^j_Rq^i_L)\,, \\
   E_1^{\text{SLL}} &= (\overline b_R\gamma_\mu\gamma_\nu q_L)(\overline b_R\gamma^\nu\gamma^\mu q_L)+8(1-\epsilon)\widetilde Q_1^{\text{SLL}}\,, \\
   E_2^{\text{SLL}} &= (\overline b^i_R\gamma_\mu\gamma_\nu q^j_L)(\overline b^j_R\gamma^\nu\gamma^\mu q^i_L)+8(1-\epsilon)Q_1^{\text{SLL}}\,.
\end{align}
The basis change between the two bases can be written in terms of two two-component vectors containing physical and evanescent operators:

\begin{equation}
  \vec{Q}=R(\vec{\overline Q} +W \vec{\overline E})\,,\qquad \vec{E}=M(\epsilon U\vec{\overline Q} +(\mathbb{1}+\epsilon UW) \vec{\overline E})\,,
\end{equation}
where the $2\times 2$ matrices $R,W,M$ and $U$ are obtained from simple tree-level relations given in \cite{Gorbahn:2009pp}. The NLO ADM of the barred basis is then translated into the second basis. In order to transform the NLO ADM from the first to the second basis a change of scheme has to be performed. It is parametrized by the quantity $Z_{QQ}^{(1,0)}$, which is a finite renormalization constant that has to be introduced to remove the finite matrix elements introduced by the physical and evanescent operators in the initial basis. This renormalization constant is given by:

\begin{equation}\label{eq:ZQQ}
  Z_{QQ}^{(1,0)}=R\left[W \overline Z^{(1,0)}_{EQ}-\left(\overline Z^{(1,1)}_{QE}+W \overline Z^{(1,1)}_{EE}-\frac{1}{2}\overline\gamma^{(0)}W\right)U\right]R^{-1}\,,
\end{equation}
which depends on the renormalization constants and the anomalous dimension matrix in the barred basis \cite{Chetyrkin:1997gb,Gorbahn:2004my}.
Adopting a similar notation as in the previous subsection we write:
\begin{equation}
  \vec{Q} = \left[R_0+\frac{\alpha_s}{4\pi}R_1\right] \vec{\overline Q}\,.
\end{equation}
In this notation the LO and NLO ADMs transform in the following way:
\begin{align}
  \gamma^{(0)} &= R_0\overline\gamma^{(0)}R_0^{-1}\,,\\
  \gamma^{(1)} &= R_0\overline\gamma^{(1)}R_0^{-1} + R_1\overline\gamma^{(0)}R_0^{-1}-R_0\overline\gamma^{(0)}R_1R_0^{-1}+2\beta_0 R_1 R_0^{-1}\,,
\end{align}
where $\beta_0$ denotes the LO $\beta$-function of the strong coupling constant.
Comparing these expressions with the results in \cite{Gorbahn:2009pp} one finds:

\begin{equation}\label{eq:R0R}
  R_0 = R\,, \quad R_1 = -Z_{QQ}^{(1,0)}R_0\,.
\end{equation}
The matrix $R_1$ therefore encodes the change of scheme between the two bases. Similar to eq.~\eqref{eq:ZQQ} it can be split into different contributions, namely:
\begin{equation}
  R_1 = R_1^{\text{shift}}+R_1^{\text{U}}\,.
\end{equation}
The first matrix $R_1^{\text{shift}}$ results from the shifts computed in the previous section and $R_1^{\text{U}}$ depends on the renormalization constants of the barred basis, which can be derived from the results in App.~\ref{app:Zs}.
\noindent
\newline
\newline
Let us first derive the matrix $R_1^{\text{shift}}$. The contribution from the Fierz-evanescent operator $\bar E_1^{\text{SLL}}$ to $R_1$ is given by the shift of $S^{LL}_{bqbq}$, which can be derived from the one of $S^{LL}_{q_1q_2q_3q_4}$ in Tab.~\ref{tab:scalar}, leading to:
\begin{equation}
  R_1^{\text{shift}}=\left(\begin{array}{cc}
    0 & 0 \\
    \frac{20+2N_c-7N_c^2}{4N_c} & \frac{4-8N_c-N_c^2}{16N_c} \\
  \end{array}\right)\,,
\end{equation}
which for $N_c=3$ corresponds to the first term in eq.~\eqref{eq:ZQQ}
\begin{equation}
  R_1^{\text{shift}}=-RWZ_{EQ}^{(1,0)}\,,
\end{equation}
where we used the relations in eq.~\eqref{eq:R0R}.
Consequently $R_1^{\text{U}}$ is given by:
\begin{equation}
  R_1^{\text{U}}=R\left(\overline Z^{(1,1)}_{QE}+W \overline Z^{(1,1)}_{EE}-\frac{1}{2}\overline\gamma^{(0)}W\right)U\,,
\end{equation}
where the relevant renormalization constants are given in App.~\ref{app:Zs}.\footnote{We use a different sign convention than \cite{Gorbahn:2009pp}.} We find
\begin{equation}
  R_1^{\text{U}}=\left(\begin{array}{cc}
  \frac{23}{6} & -\frac{1}{24} \\
\frac{43}{12} & \frac{19}{48} \\
  \end{array}\right)\,,
\end{equation}
leading to
\begin{equation}
  Z_{QQ}^{(1,0)} = -(R_1^{\text{shift}}+R_1^{\text{U}})R_0^{-1}=\left(\begin{array}{cc}
  -\frac{11}{3} & \frac{1}{3} \\
 \frac{1}{3} & \frac{5}{3} \\
  \end{array}\right)\,,
\end{equation}
which agrees with the findings of \cite{Gorbahn:2009pp}. This simple exercise shows that if the matrix $U$ is non-zero, which corresponds to physical operators mixing into the evanescent operators in the basis change, the transformation is slightly more involved and besides the one-loop shifts one also needs to take into account the renormalization constants given in App.~\ref{app:Zs}.

\section{Conclusions}\label{sec:conclusions}

We present for the first time one-loop Fierz transformations that include QCD and QED corrections, which result from Fierz-evanescent operators. The corrections are computed in a general renormalisation scheme, that can easily be transformed into the original BMU scheme, in which most of the two-loop ADMs are known.

\noindent
Our results simplify one-loop matching calculations in which Fierz transformations are needed, since only the matching of physical operators has to be computed, whereas the evanescent part is given by our formulae. They therefore facilitate the projection onto the physical basis and might for example be relevant for common matching tools such as \texttt{matchmakereft} \cite{Carmona:2021xtq}.

\noindent
A second use case of our results are basis transformations involving two-loop ADMs. Changing a two-loop ADM from one basis into another is particularly simple in the absence of physical operators mixing into evanescent ones in the transformation. In that case the change of scheme can be read of directly from the one-loop Fierz relations. A general one-loop basis transformation can however be performed using the one-loop Fierz relations in combination with the renormalisation constants for the Fierz-evanescent operators provided in the Appendix.
One-loop basis changes of two-loop ADMs play for instance an important role in the NLO SMEFT analysis, since in that case the one-loop matching is known in the JMS basis whereas the two-loop running is given in the BMU basis. The results of this article allow to perform the full JMS-BMU translation at one-loop, which paves the way for a complete and scheme-independent NLO SMEFT analysis. For that purpose we plan to implement our results in the common SMEFT computing tools like \texttt{WCxf} \cite{Aebischer:2017ugx} and \texttt{wilson} \cite{Aebischer:2018bkb} and finally in the basis change package \texttt{abc\_eft} \cite{Proceedings:2019rnh}.

\noindent
As a possible extension of our results the computations can be performed for a general gauge group. Furthermore Yukawa one-loop corrections to the tree-level Fierz identities could be considered, which we will leave for the future.


\section*{Acknowledgements}
\addcontentsline{toc}{section}{\numberline{}Acknowledgements}

We thank Gino Isidori and Andrzej Buras for comments on the manuscript. J.\ A.\ and M.\ P.\ acknowledge  financial  support  from  the  European  Research  Council  (ERC)  under the European Union's Horizon 2020 research and innovation programme under grant agreement 833280 (FLAY), and from the Swiss National Science Foundation (SNF) under contract 200020-204428.


\clearpage

\appendix

\newpage

\section{Generalised Greek Identities}\label{app:greek}
In this Appendix we report the generalised Greek identities used in our calculations. The original Greek identities used in the BMU scheme can be found in the Appendix of \cite{Buras:2012fs} and are obtained from the generalised version by setting all constants to one, $a_{1,2,3}=b_{1,2,3}=c_{1,2,3}=d_{1,2,3}=e_{1,2}=f_{1,2}=1$. Another difference is the definition of $\sigma^{\mu \nu} = \frac{i}{2}\left[ \gamma^\mu, \gamma^\nu \right]$, which introduces a negative sign for the tensor contributions.

\vspace{0.3cm}
\underline{VLL}
\begin{align}
  \gamma_{\alpha} \gamma_{\beta} \gamma_{\mu}\left(1 \pm \gamma_{5}\right) \gamma^{\beta} \gamma^{\alpha} \otimes \gamma^{\mu}\left(1 \pm \gamma_{5}\right)&=4(1-2a_1 \epsilon) \gamma_{\mu}\left(1 \pm \gamma_{5}\right) \otimes \gamma^{\mu}\left(1 \pm \gamma_{5}\right)\,, \\
  \gamma_{\mu}\left(1 \pm \gamma_{5}\right) \gamma_{\alpha} \gamma_{\beta} \otimes \gamma^{\mu}\left(1 \pm \gamma_{5}\right) \gamma^{\alpha} \gamma^{\beta}&=4(4-a_2\epsilon) \gamma_{\mu}\left(1 \pm \gamma_{5}\right) \otimes \gamma^{\mu}\left(1 \pm \gamma_{5}\right)\,, \\
  \gamma_{\mu}\left(1 \pm \gamma_{5}\right) \gamma_{\alpha} \gamma_{\beta} \otimes \gamma^{\beta} \gamma^{\alpha} \gamma^{\mu}\left(1 \pm \gamma_{5}\right)&=4(1-2 a_3\epsilon) \gamma_{\mu}\left(1 \pm \gamma_{5}\right) \otimes \gamma^{\mu}\left(1 \pm \gamma_{5}\right)\,.
\end{align}

\underline{VLR}
\begin{align}
  \gamma_{\alpha} \gamma_{\beta} \gamma_{\mu}\left(1 \pm \gamma_{5}\right) \gamma^{\beta} \gamma^{\alpha} \otimes \gamma^{\mu}\left(1 \mp \gamma_{5}\right)&=4(1-2b_1 \epsilon) \gamma_{\mu}\left(1 \pm \gamma_{5}\right) \otimes \gamma^{\mu}\left(1 \mp \gamma_{5}\right)\,,
  \\
  \gamma_{\mu}\left(1 \pm \gamma_{5}\right) \gamma_{\alpha} \gamma_{\beta} \otimes \gamma^{\mu}\left(1 \mp \gamma_{5}\right) \gamma^{\alpha} \gamma^{\beta}&=4(1+b_2\epsilon) \gamma_{\mu}\left(1 \pm \gamma_{5}\right) \otimes \gamma^{\mu}\left(1 \mp \gamma_{5}\right)\,, \\
  \gamma_{\mu}\left(1 \pm \gamma_{5}\right) \gamma_{\alpha} \gamma_{\beta} \otimes \gamma^{\beta} \gamma^{\alpha} \gamma^{\mu}\left(1 \mp \gamma_{5}\right)&=16(1-b_3\epsilon) \gamma_{\mu}\left(1 \pm \gamma_{5}\right) \otimes \gamma_{\mu}\left(1 \mp \gamma_{5}\right)\,.
\end{align}

\underline{SLR}
\begin{align}
  \gamma_{\nu} \gamma_{\mu}\left(1 \mp \gamma_{5}\right) \gamma^{\mu} \gamma^{\nu} \otimes\left(1 \pm \gamma_{5}\right)&=16(1-c_1\epsilon)\left(1 \mp \gamma_{5}\right) \otimes\left(1 \pm \gamma_{5}\right)\,, \\
  \left(1 \mp \gamma_{5}\right) \gamma_{\mu} \gamma_{\nu} \otimes\left(1 \pm \gamma_{5}\right) \gamma^{\mu} \gamma^{\nu}&=4(1+c_2\epsilon)\left(1 \mp \gamma_{5}\right) \otimes\left(1 \pm \gamma_{5}\right)\,, \\
  \left(1 \mp \gamma_{5}\right) \gamma_{\nu} \gamma_{\mu} \otimes \gamma^{\mu} \gamma^{\nu}\left(1 \pm \gamma_{5}\right)&=4(1-2 c_3\epsilon)\left(1 \mp \gamma_{5}\right) \otimes\left(1 \pm \gamma_{5}\right)\,.
\end{align}

\underline{SLL}
\begin{align}
&\gamma_{\nu} \gamma_{\mu}\left(1 \pm \gamma_{5}\right) \gamma^{\mu} \gamma^{\nu} \otimes\left(1 \pm \gamma_{5}\right)=16(1-d_1\epsilon)\left(1 \pm \gamma_{5}\right) \otimes\left(1 \pm \gamma_{5}\right)\,, \\
&\left(1 \pm \gamma_{5}\right) \gamma_{\mu} \gamma_{\nu} \otimes\left(1 \pm \gamma_{5}\right) \gamma^{\mu} \gamma^{\nu}\notag\\
&=(4-2 d_2\epsilon)\left(1 \pm \gamma_{5}\right) \otimes\left(1 \pm \gamma_{5}\right)
-\sigma_{\mu \nu}\left(1 \pm \gamma_{5}\right) \otimes \sigma^{\mu \nu}\left(1 \pm \gamma_{5}\right)\,, \\
&\left(1 \pm \gamma_{5}\right) \gamma_{\mu} \gamma_{\nu} \otimes \gamma^{\nu} \gamma^{\mu}\left(1 \pm \gamma_{5}\right) \notag\\
&=(4-2 d_3\epsilon)\left(1 \pm \gamma_{5}\right) \otimes\left(1 \pm \gamma_{5}\right)
+\sigma_{\mu \nu}\left(1 \pm \gamma_{5}\right) \otimes \sigma^{\mu \nu}\left(1 \pm \gamma_{5}\right)\,.
\end{align}

\underline{TLL}
\begin{align}
&\gamma^{\alpha} \gamma^{\beta} \sigma^{\mu \nu}\left(1 \pm \gamma_{5}\right) \gamma_{\beta} \gamma_{\alpha} \otimes \sigma^{\mu \nu}\left(1 \pm \gamma_{5}\right)=0 \,,\\
&\sigma_{\mu \nu}\left(1 \pm \gamma_{5}\right) \gamma_{\alpha} \gamma_{\beta} \otimes \sigma^{\mu \nu}\left(1 \pm \gamma_{5}\right) \gamma^{\alpha} \gamma^{\beta} \notag\\
&\quad=-(48-80 e_1\epsilon)\left(1 \pm \gamma_{5}\right) \otimes\left(1 \pm \gamma_{5}\right)+(12-6 f_1\epsilon) \sigma_{\mu \nu}\left(1 \pm \gamma_{5}\right) \otimes \sigma^{\mu \nu}\left(1 \pm \gamma_{5}\right)\,, \\
&\sigma_{\mu \nu}\left(1 \pm \gamma_{5}\right) \gamma_{\alpha} \gamma_{\beta} \otimes \gamma^{\beta} \gamma^{\alpha} \sigma^{\mu \nu}\left(1 \pm \gamma_{5}\right) \notag\\
&\quad=(48-80 e_2\epsilon)\left(1 \pm \gamma_{5}\right) \otimes\left(1 \pm \gamma_{5}\right)+(12-14 f_2\epsilon) \sigma_{\mu \nu}\left(1 \pm \gamma_{5}\right) \otimes \sigma^{\mu \nu}\left(1 \pm \gamma_{5}\right)\,.
\end{align}
Note that we introduce the same evanescent structures for operators with colour-crossed $SU(3)_c$ structure as well as for Parity-flipped operators.

\section{Loop Structures}\label{app:loop}
In this Appendix we report the loop structures that are encountered in our calculation. If another scheme is used, these structures can be used to compute the shifts in the new scheme. We list the "Feynman-rules" for the divergent parts of these structures in QCD and QED below.

\subsection{QCD}
We consider the vertex corrections of a four-quark operator with the following flavour content
\begin{equation}
  \mathcal{O} = (\overline q_1 \Gamma_1 q_2)(\overline q_3 \Gamma_2 q_4)\,.
\end{equation}
To be general we suppress the colour contraction of the operator $\mathcal{O}$, which then has to be added to the "Feynman-rules" listed below. We only list three out of the six vertex corrections, since the poles of the other three amplitudes are identical to the ones shown.
\newline
\newline

\begin{tabular}{
    >{\raggedright\arraybackslash} m{0.4\textwidth}
    >{\raggedright\arraybackslash} m{0.2\textwidth}}
\hspace*{-0.7cm}
  \includegraphics[width=0.45\textwidth]{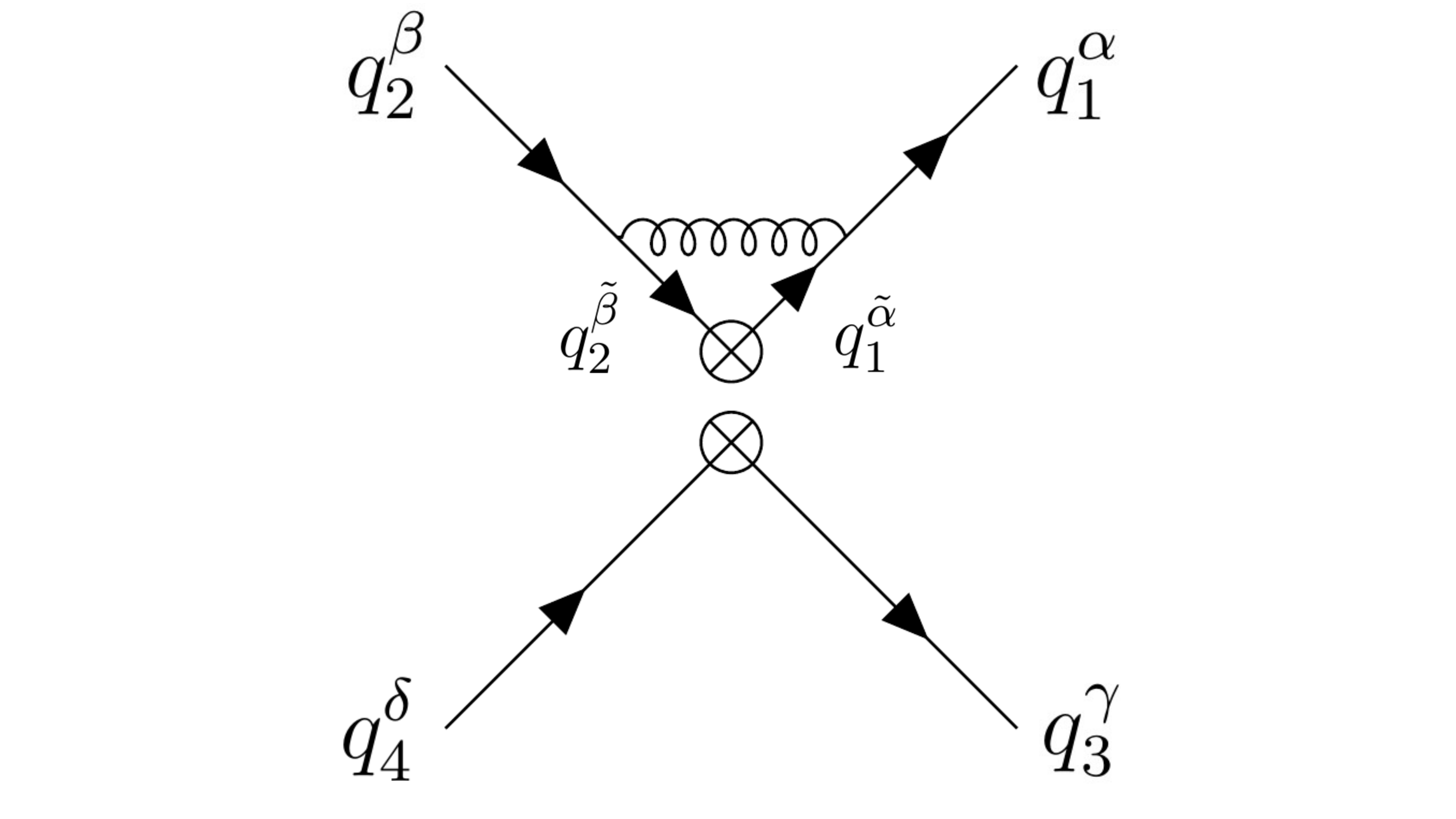} &
  $
  \displaystyle
\hspace*{+0.1cm}
  \frac{\alpha_s}{4\pi}\frac{1}{4\epsilon}(T^A)_{\alpha \widetilde\alpha}(T^A)_{\widetilde\beta \beta }(\overline q_1^\alpha \gamma_\mu\gamma_\nu \Gamma_1 \gamma^\mu\gamma^\nu q_2^\beta)(\overline q_3^\gamma \Gamma_2 q_4^\delta)
  $ \\
\hspace*{-0.7cm}
  \includegraphics[width=0.45\textwidth]{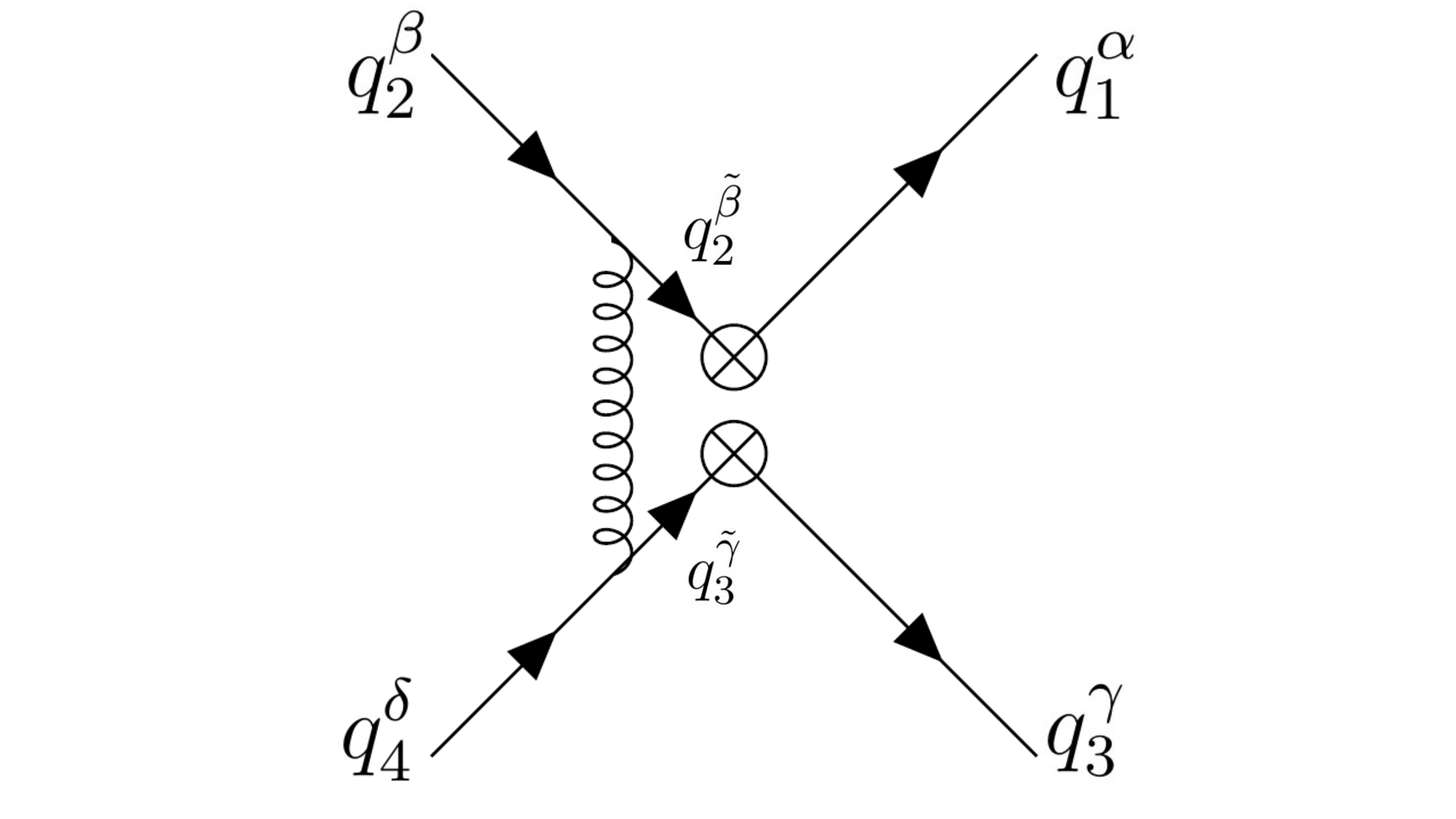} &
  $
  \displaystyle
\hspace*{+0.1cm}
  -\frac{\alpha_s}{4\pi}\frac{1}{4\epsilon}(T^A)_{\widetilde\beta \beta}(T^A)_{\widetilde\gamma \delta}(\overline q_1^\alpha  \Gamma_1\gamma_\mu\gamma_\nu q_2^\beta)(\overline q_3^\gamma \Gamma_2 \gamma^\mu\gamma^\nu q_4^\delta)
  $ \\
 \end{tabular}

\begin{tabular}{
    >{\raggedright\arraybackslash} m{0.4\textwidth}
    >{\raggedright\arraybackslash} m{0.2\textwidth}}
\hspace*{-0.7cm}
  \includegraphics[width=0.45\textwidth]{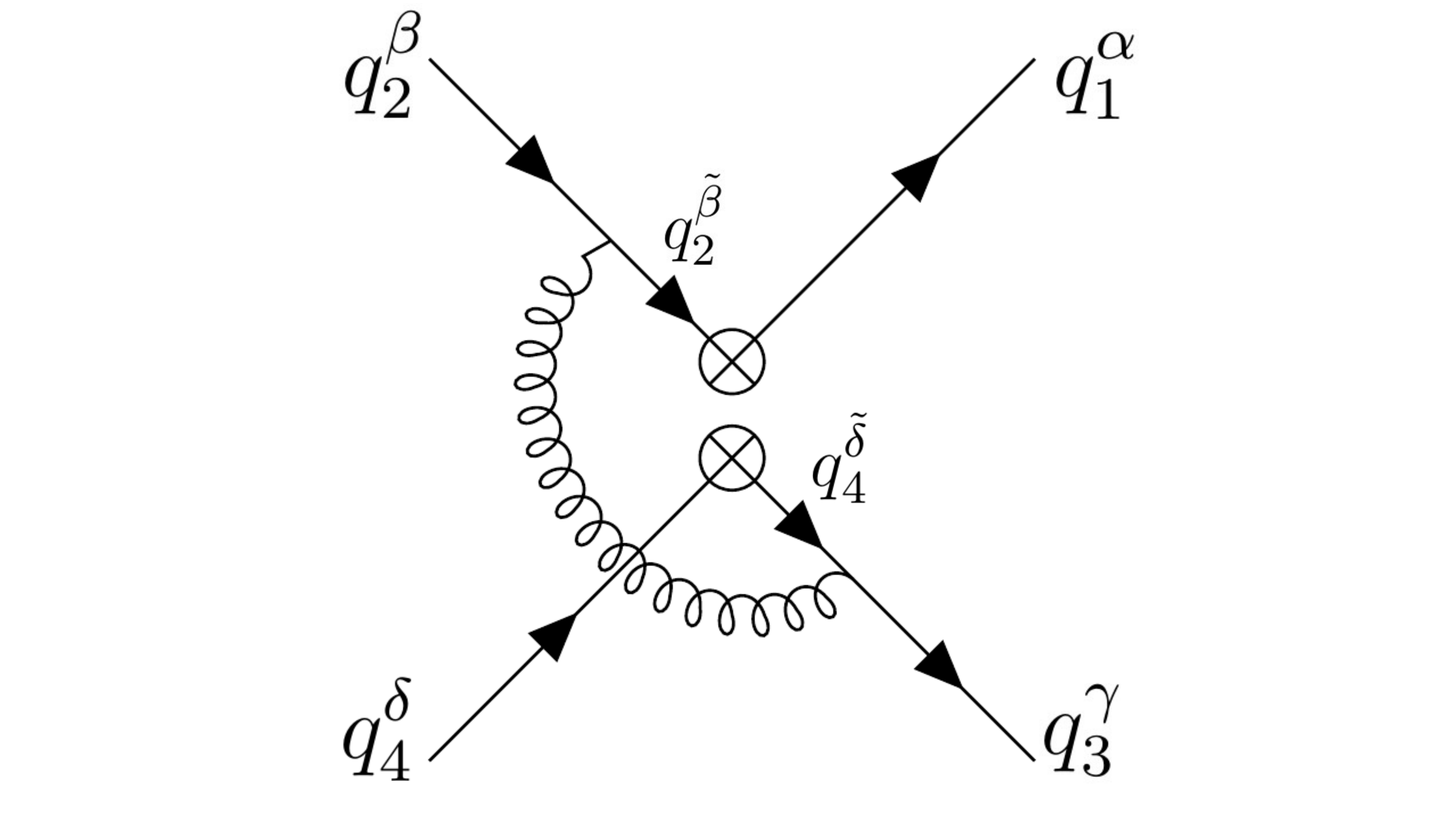} &
  $
  \displaystyle
\hspace*{+0.1cm}
  \frac{\alpha_s}{4\pi}\frac{1}{4\epsilon}(T^A)_{\widetilde\beta \beta}(T^A)_{\gamma\widetilde\delta}(\overline q_1^\alpha  \Gamma_1\gamma_\mu\gamma_\nu q_2^\beta)(\overline q_3^\gamma\gamma^\nu\gamma^\mu  \Gamma_2 q_4^\delta)
  $ \\
 \end{tabular}

\noindent
For the penguin contributions we consider the operator
\begin{equation}
  \mathcal{P} = (\overline q_1 \Gamma_1 q_2)(\overline q_3 \Gamma_2 q_3)\,,
\end{equation}
where we again suppress the colour contraction, which has to be added to the "Feynman-rules" below. One finds:
\newline
\newline

\begin{tabular}{
    >{\raggedright\arraybackslash} m{0.5\textwidth}
    >{\raggedright\arraybackslash} m{5\textwidth}}
\hspace*{-2.25cm}
  \includegraphics[width=0.65\textwidth]{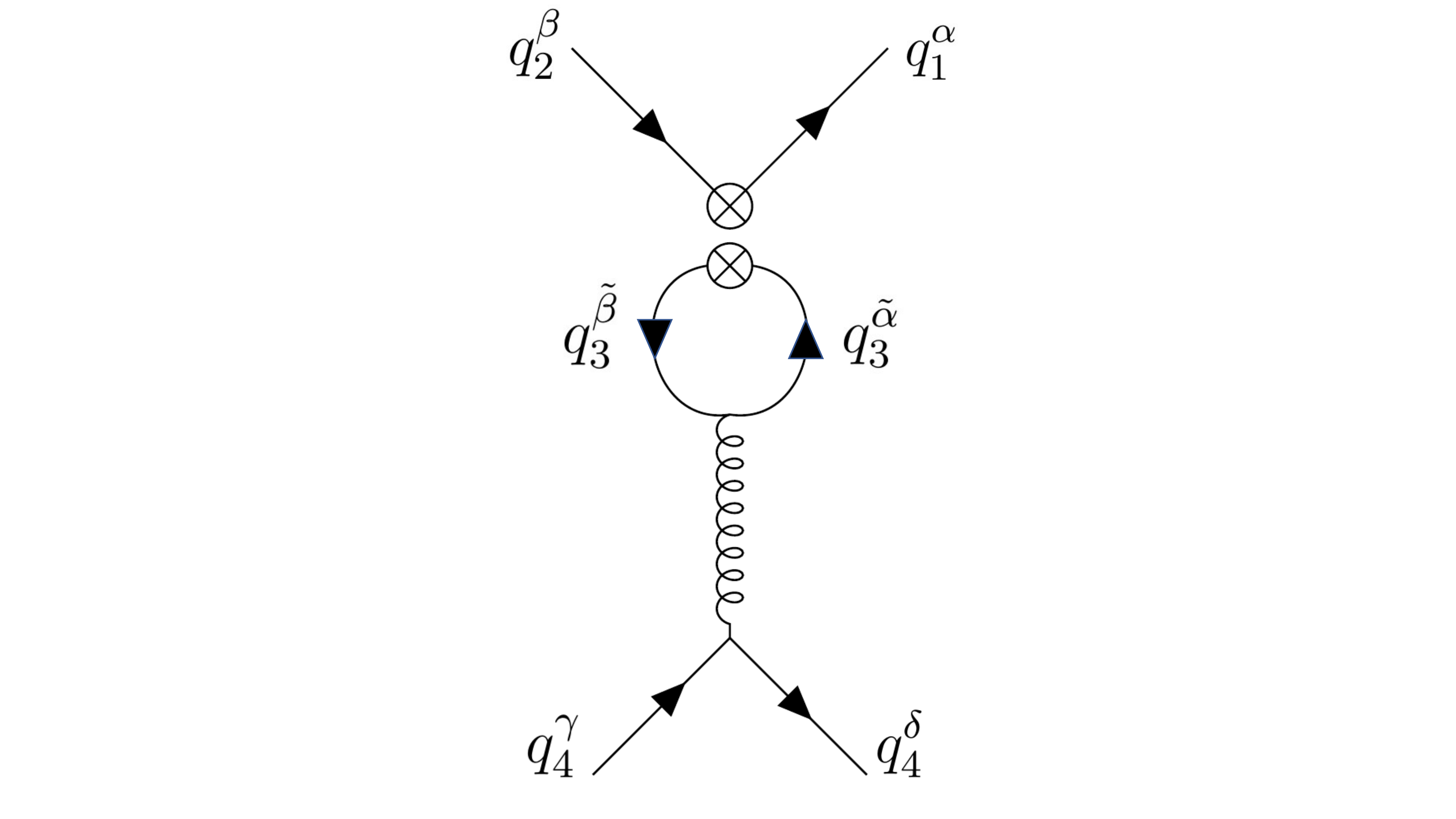} &
  $
  \displaystyle
\hspace*{-1.3cm}
  -\frac{\alpha_s}{4\pi}\frac{D}{12\epsilon}(T^A)_{\delta\gamma}(T^A)_{\widetilde\delta\widetilde\gamma}(\overline q_1^\alpha  \Gamma_1 q_2^\beta)(\overline q_4^{\widetilde\beta}\gamma^\mu q_4^{\widetilde\delta})\text{Tr}[\gamma_\mu \Gamma_2]
  $ \\
\hspace*{-2.1cm}
  \includegraphics[width=0.65\textwidth]{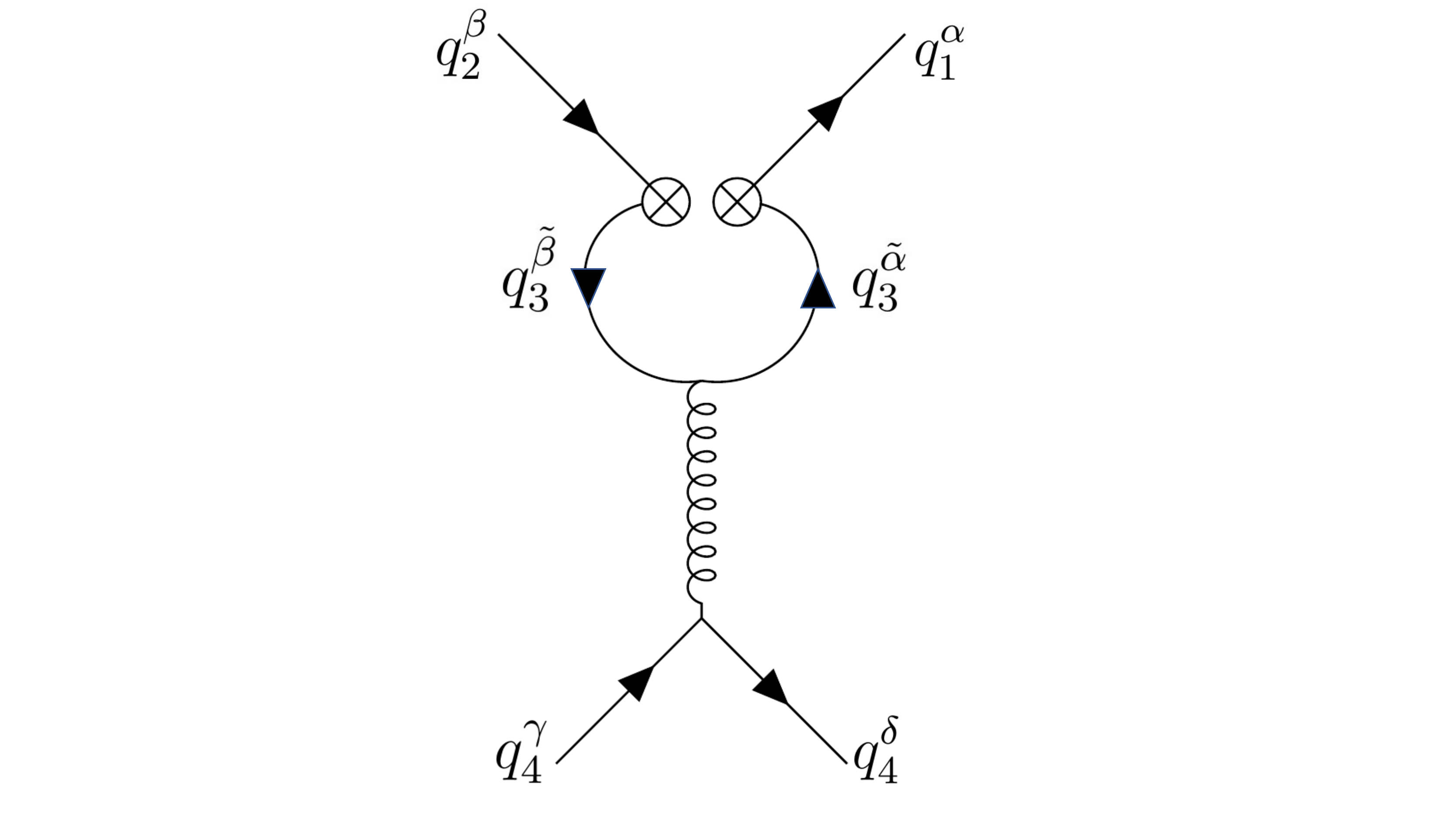} &
  $
  \displaystyle
\hspace*{-1.3cm}
  \frac{\alpha_s}{4\pi}\frac{D}{12\epsilon}(T^A)_{\delta\gamma}(T^A)_{\widetilde\delta\widetilde\gamma}(\overline q_1^\alpha  \Gamma_1 \gamma_\mu\Gamma_2 q_2^\beta)(\overline q_4^{\widetilde\beta}\gamma^\mu q_4^{\widetilde\delta})
  $ \\
 \end{tabular}

\newpage
\subsection{QED}

For the vertex corrections we consider the operator

\begin{equation}
  \mathcal{O}_{\text{QED}} = (\overline f_1 \Gamma_1 f_2)(\overline f_3 \Gamma_2 f_4)\,,
\end{equation}
\noindent
with four different fermions. We find:
\newline
\newline

\begin{tabular}{
    >{\raggedright\arraybackslash} m{0.5\textwidth}
    >{\raggedright\arraybackslash} m{5\textwidth}}
\hspace*{-0.7cm}
  \includegraphics[width=0.45\textwidth]{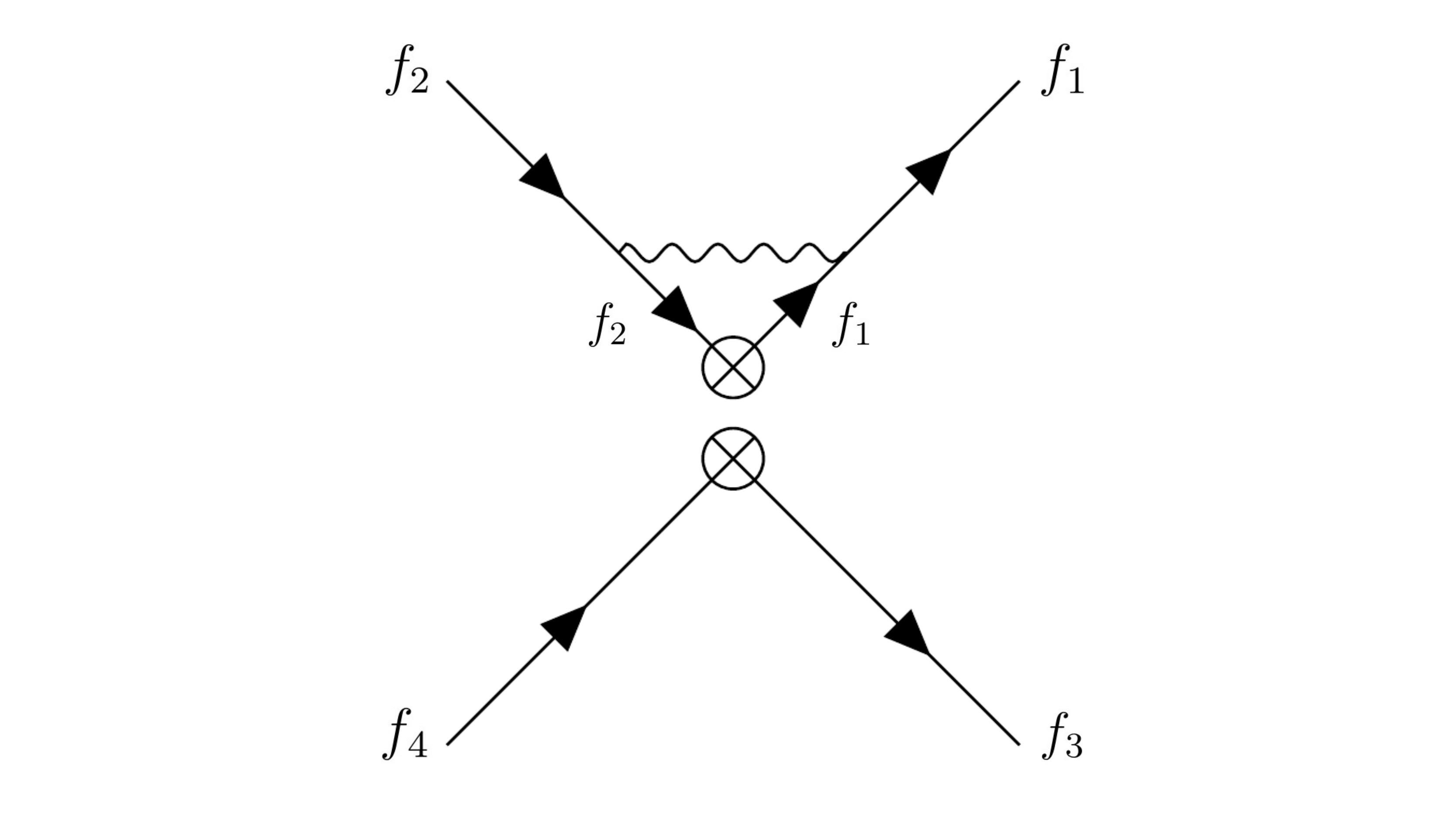} &
  $
  \displaystyle
\hspace*{-1.4cm}
  \frac{\alpha_e}{4\pi}Q_1Q_2\frac{1}{4\epsilon}(\overline f_1 \gamma_\mu\gamma_\nu \Gamma_1 \gamma^\mu\gamma^\nu f_2)(\overline f_3 \Gamma_2 f_4)
  $ \\
\hspace*{-0.7cm}
  \includegraphics[width=0.45\textwidth]{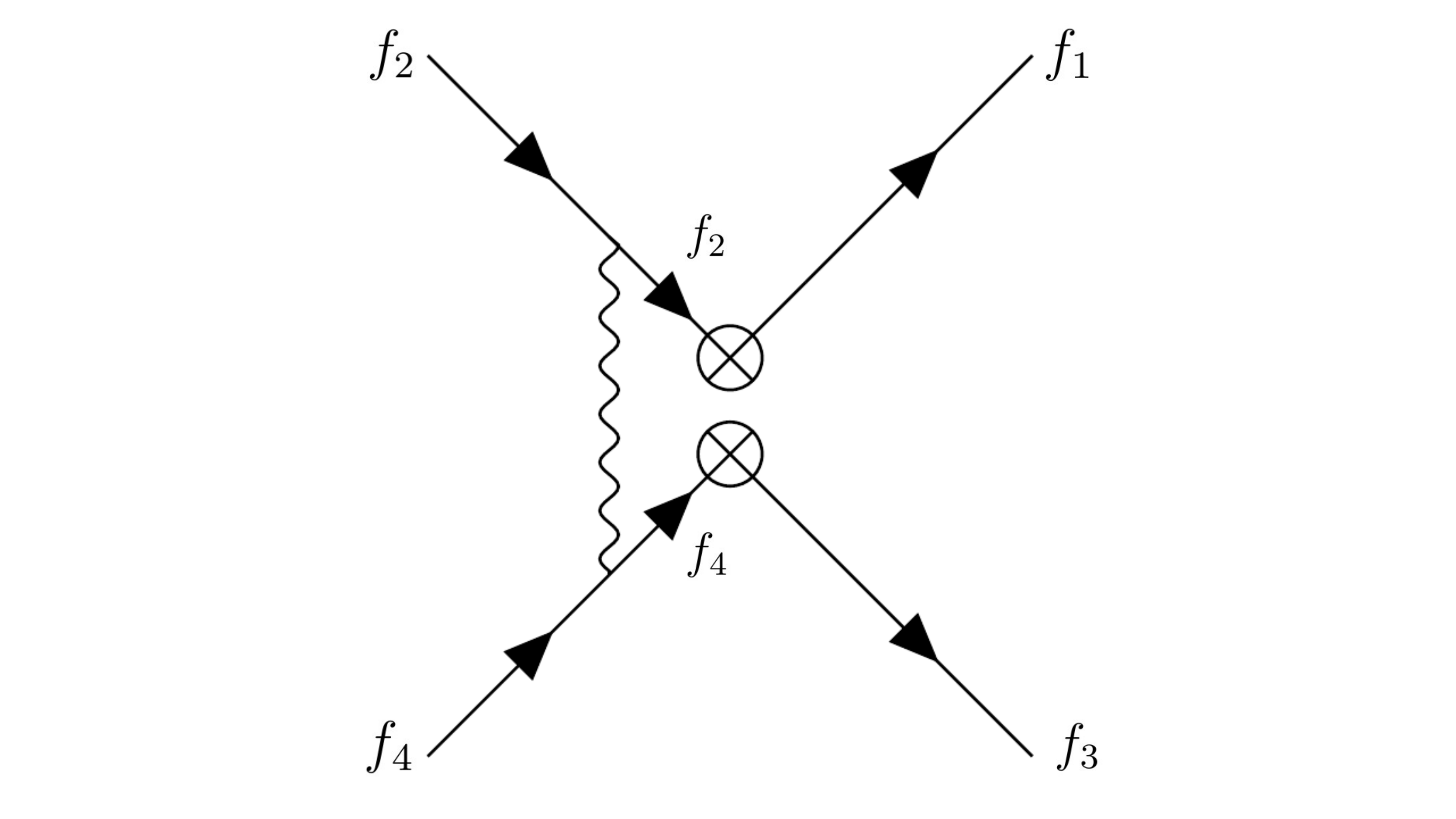} &
  $
  \displaystyle
\hspace*{-1.4cm}
  -\frac{\alpha_e}{4\pi}Q_2Q_4\frac{1}{4\epsilon}(\overline f_1  \Gamma_1 \gamma_\mu\gamma_\nu f_2)(\overline f_3 \Gamma_2\gamma^\mu\gamma^\nu f_4)
  $ \\
\hspace*{-0.7cm}
  \includegraphics[width=0.45\textwidth]{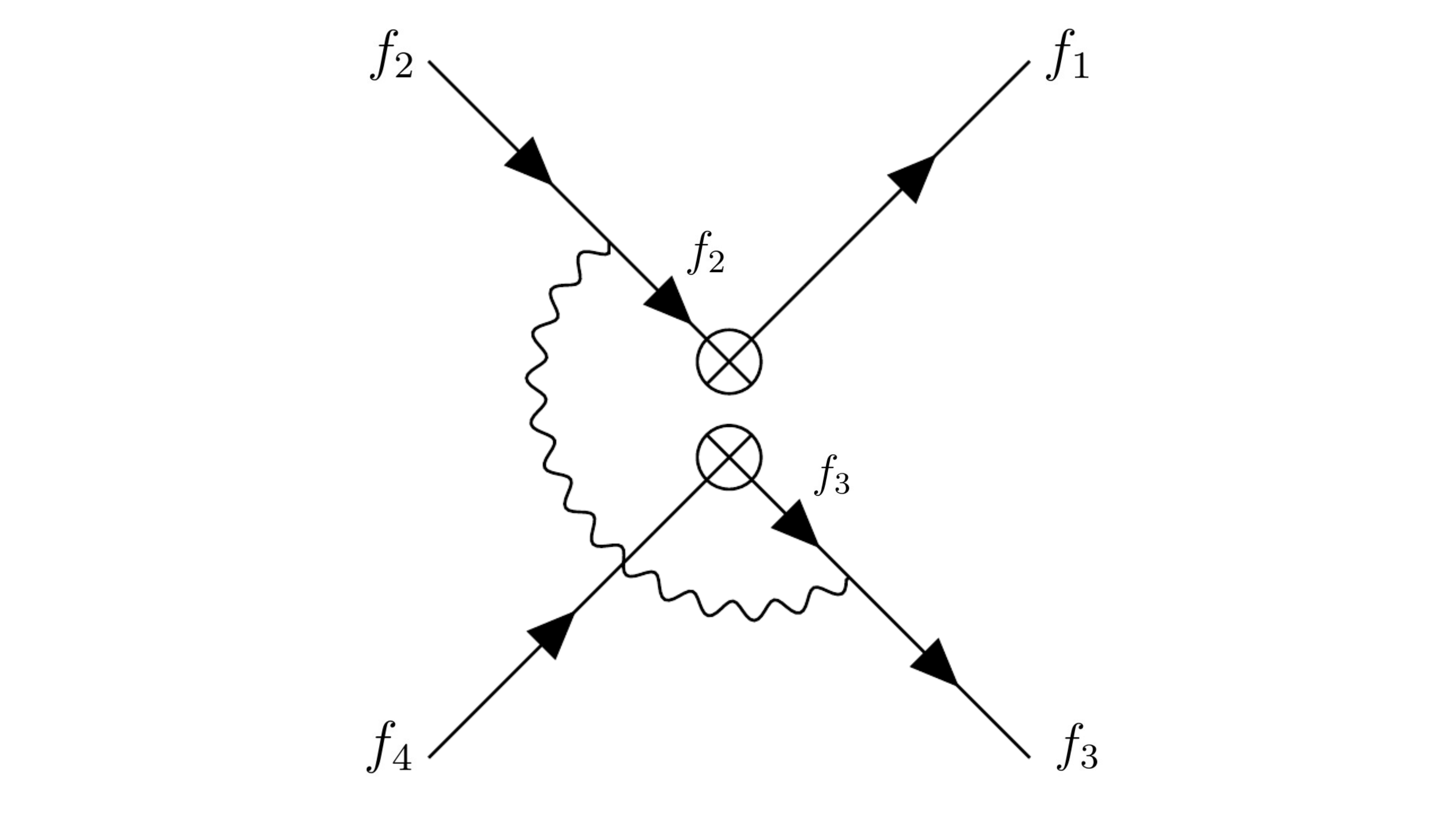} &
  $
  \displaystyle
\hspace*{-1.4cm}
  \frac{\alpha_e}{4\pi}Q_2Q_3\frac{1}{4\epsilon}(\overline f_1  \Gamma_1 \gamma_\mu\gamma_\nu f_2)(\overline f_3 \gamma^\nu\gamma^\mu\Gamma_2 f_4)
  $ \\
 \end{tabular}
\newpage
\noindent
 For the penguin contributions we consider insertions of the operator

 \begin{equation}
   \mathcal{P}_{\text{QED}} = (\overline f_1 \Gamma_1 f_2)(\overline f_3 \Gamma_2 f_3)\,,
 \end{equation}
\noindent
For closed and open penguins we find the following divergent structures:
\newline
\newline

 \begin{tabular}{
     >{\raggedright\arraybackslash} m{0.5\textwidth}
     >{\raggedright\arraybackslash} m{5\textwidth}}
\hspace*{-2.1cm}
   \includegraphics[width=0.65\textwidth]{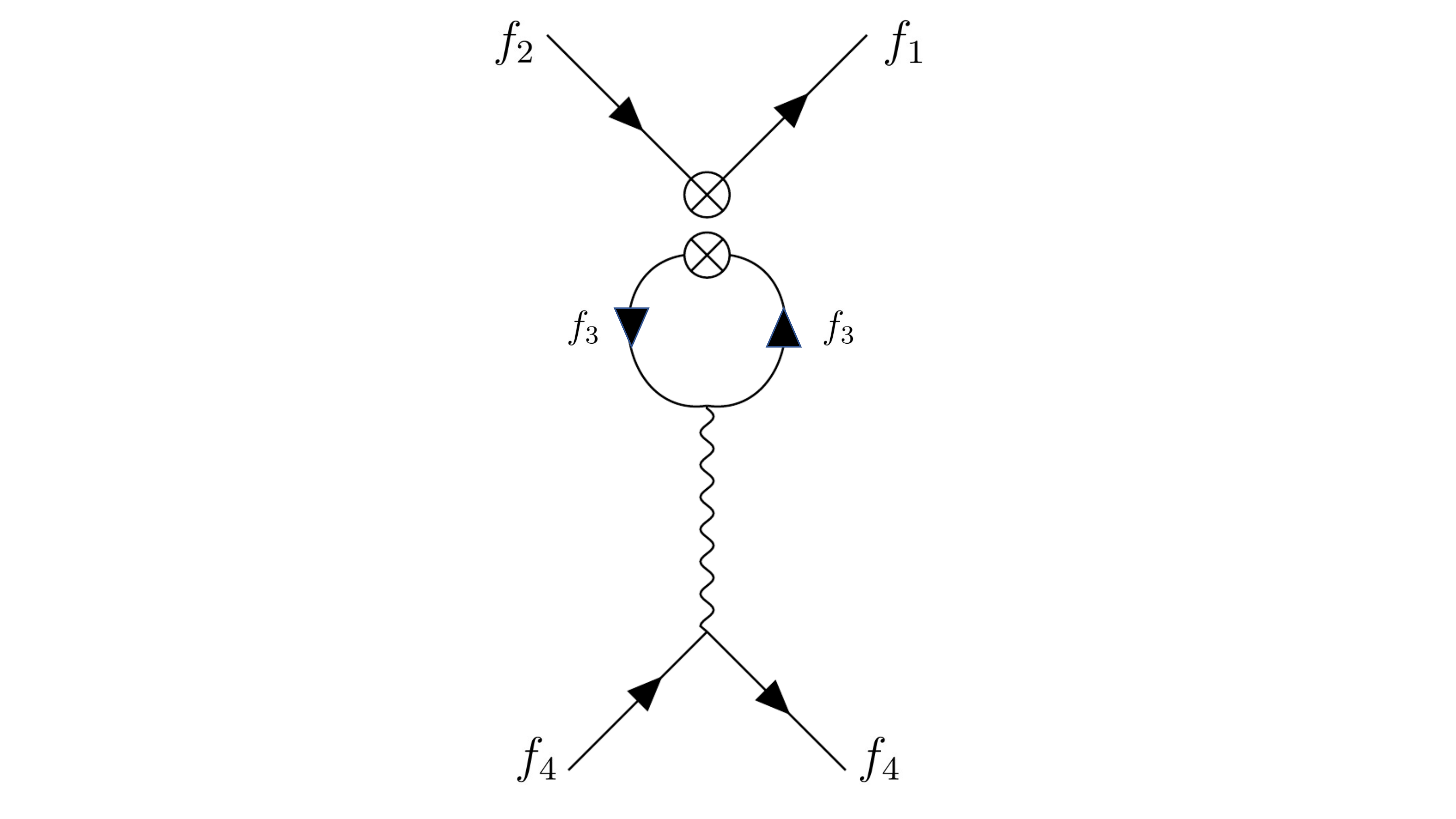} &
   $
   \displaystyle
\hspace*{-1.4cm}
     -\frac{\alpha_e}{4\pi}\frac{D}{12\epsilon}Q_3^2(\overline f_1  \Gamma_1 f_2)(\overline q_4\gamma^\mu f_4)\text{Tr}[\gamma_\mu \Gamma_2]
   $ \\
\hspace*{-2cm}
   \includegraphics[width=0.65\textwidth]{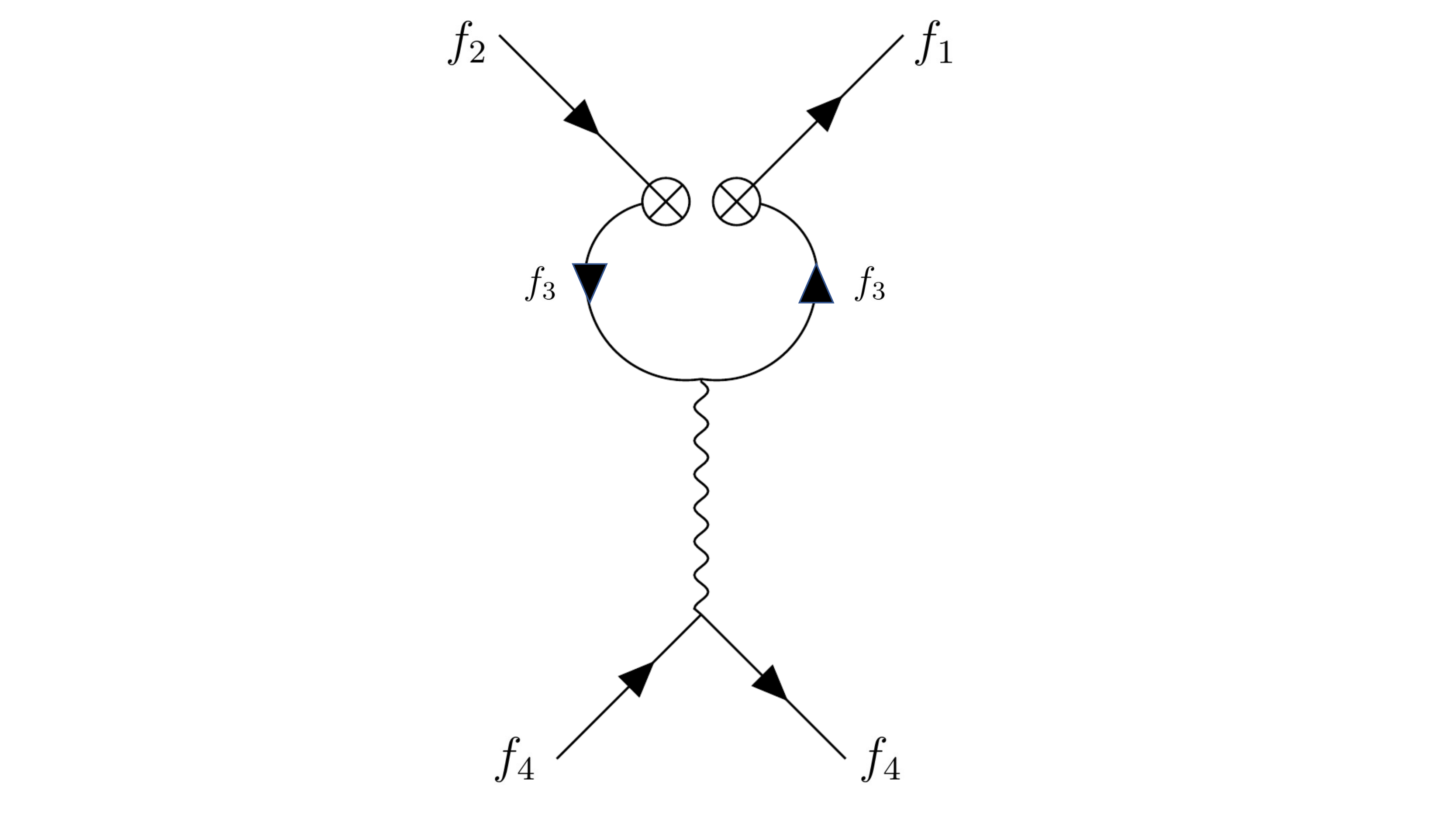} &
   $
   \displaystyle
\hspace*{-1.4cm}
   \frac{\alpha_s}{4\pi}\frac{D}{12\epsilon}Q_3^2(\overline f_1  \Gamma_1 \gamma_\mu\Gamma_2 f_2)(\overline f_4\gamma^\mu f_4)
   $ \\
  \end{tabular}

\newpage
\section{General Shifts}\label{app:genshifts}

In this appendix we report the one-loop shifts that result from the generalised BMU scheme. The results for VLL operators are given in Tab.~\ref{tab:genBMUVLL}.

\begin{table}[H]
\renewcommand{\arraystretch}{1.5}
\small
\begin{adjustwidth}{-0.5cm}{}
\begin{align*}
\begin{array}[t]{c|c|c|c}
\toprule
 \text{Operator} & \text{Tree-level Fierz} & \text{QCD shift} &  \text{QED shift}   \\
\midrule\midrule
\op{V}{LL}{q_1q_2q_3q_4} & \cop{V}{LL}{q_1q_4q_3q_2} & 2(a_1-a_3)\left( \cop{V}{LL}{q_1q_2q_3q_4} -N_c\op{V}{LL}{q_1q_2q_3q_4}\right) & -2(a_1-a_3) \left(Q_{1234}-Q_{1423}\right) \op{V}{LL}{q_1q_2q_3q_4}    \\
\cop{V}{LL}{q_1q_2q_3q_4} & \op{V}{LL}{q_1q_4q_3q_2} & 2(a_3-a_1)\left( \op{V}{LL}{q_1q_2q_3q_4} -N_c\cop{V}{LL}{q_1q_2q_3q_4}\right) & -2 (a_1-a_3)\left(Q_{1234}-Q_{1423}\right) \cop{V}{LL}{q_1q_2q_3q_4}  \\
\op{V}{LL}{q_1q_2\ell_1\ell_2} & \op{V}{LL}{q_1\ell_2\ell_1q_2} & (a_1-a_3)\frac{1-N_c^2}{N_c}\op{V}{LL}{q_1q_2\ell_1\ell_2}  & -2 (a_1-a_3)\left(Q_{1234}-Q_{1423}\right) \op{V}{LL}{q_1q_2\ell_1\ell_2} \\
\op{V}{LL}{q_1\ell_1\ell_2q_2} & \op{V}{LL}{q_1q_2\ell_2\ell_1} &  (a_3-a_1)\frac{1-N_c^2}{N_c}\op{V}{LL}{q_1\ell_1\ell_2q_2}  & -2 (a_1-a_3)\left(Q_{1234}-Q_{1423}\right) \op{V}{LL}{q_1\ell_1\ell_2q_2} \\
\op{V}{LL}{\ell_1\ell_2\ell_3\ell_4} & \op{V}{LL}{\ell_1\ell_4\ell_3\ell_2} & 0  & -2 (a_1-a_3)\left(Q_{1234}-Q_{1423}\right) \op{V}{LL}{\ell_1\ell_2\ell_3\ell_4} \\
\bottomrule
\end{array}
\end{align*}
\end{adjustwidth}
\caption{Fierz transformations for four-quark $\gamma_\mu P_L \otimes \gamma^\mu P_L$ operators together with their one-loop QCD and QED shifts resulting from genuine vertex corrections in the generalised BMU scheme.}
\label{tab:genBMUVLL}
\end{table}
\noindent
The shifts for SLR and VRL operators are non-vanishing in the generalised BMU scheme, which can be seen in Tab.~\ref{tab:genBMULR}. For the QED corrections to these operators we define the quantity

\begin{equation}
  A_{vwxy} = \left[2 Q_{1423} (b_1-c_3)+Q_{1324} (b_2-c_2)+4Q_{1234} (b_3-c_1)\right]\op{S}{LR}{vwxy}\,.
\end{equation}

\begin{table}[H]
\begin{adjustwidth}{0cm}{}
\renewcommand{\arraystretch}{1.5}
\small
\begin{align*}
\begin{array}[t]{c|c|c|c}
\toprule
 \text{Operator} & \text{Tree-level Fierz} & \text{QCD shift} &  \text{QED shift}   \\
\midrule\midrule
\op{S}{LR}{q_1q_2q_3q_4} &
-\frac{1}{2} \cop{V}{RL}{q_1q_4q_3q_2} &
\makecell{ \left(\frac{-2b_1-b_2+c_2+2c_3}{N_c}+4\left(b_3-c_1 \right)\frac{N_c^2-1}{N_c}\right)\op{S}{LR}{q_1q_2q_3q_4}\\+(2b_1+b_2-c_2-2c_3)\cop{S}{LR}{q_1q_2q_3q_4}}
& A_{q_1q_2q_3q_4}  \\
& & & \\
\cop{S}{LR}{q_1q_2q_3q_4}
&-\frac{1}{2} \op{V}{RL}{q_1q_4q_3q_2}
&\makecell{(b_2+4b_3-4c_1-c_2)\op{S}{LR}{q_1q_2q_3q_4}\\
+ \left(\frac{-2b_1-b_2-4b_3+4c_1+c_2+2c_3}{N_c}+2\left(b_1-c_3 \right)N_c\right)\cop{S}{LR}{q_1q_2q_3q_4}}
& \widetilde A_{q_1q_2q_3q_4} \\
& & & \\
\op{S}{LR}{q_1q_2\ell_1\ell_2}
&-\frac{1}{2} \op{V}{RL}{q_1\ell_2\ell_1q_2}
&\makecell{2(b_3-c_1)\frac{N_c^2-1}{N_c}\op{S}{LR}{q_1q_2\ell_1\ell_2}}
& A_{q_1q_2\ell_1\ell_2}\\
& & & \\
\op{S}{LR}{q_1\ell_2\ell_1q_2}
&-\frac{1}{2} \op{V}{RL}{q_1q_2\ell_1\ell_2}
&\makecell{(b_1-c_3)\frac{N_c^2-1}{N_c}\op{S}{LR}{q_1\ell_2\ell_1q_2}}
& A_{q_1\ell_2\ell_1q_2}\\
& & & \\
\op{S}{LR}{\ell_1\ell_2\ell_3\ell_4}
&-\frac{1}{2} \op{V}{RL}{\ell_1\ell_4\ell_3\ell_2}
&\makecell{0}
&A_{\ell_1\ell_2\ell_3\ell_4}\\
\bottomrule
\end{array}
\end{align*}
\end{adjustwidth}
\caption{Fierz transformations for scalar $P_L \otimes P_R$ operators together with their one-loop QCD and QED shifts in the generalised BMU scheme.}
\label{tab:genBMULR}
\end{table}
\noindent
The SLL operators and their shifts in the generalised BMU scheme are collected in Tab.~\ref{tab:SLLgenbmu}. For the QED shifts to these operators we define the quantity:
\begin{align}
&B_{vwxy}\equiv \nonumber\\
&\frac{1}{32}\Big[\left(-128d_1+4d_3+40e_2+84f_2\right)Q_{1234}+\left(12d_2+40e_1-36f_1\right)Q_{1324}\nonumber\\& \,\,\, +\left(32d_1-16d_3\right)Q_{1423} \Big]\op{S}{LL}{vwxy}\nonumber\\
&\frac{1}{32}\Big[\left(d_3+10e_2-7f_2\right)Q_{1234}+\left(-d_2+10e_1+3f_1\right)Q_{1324}+8d_1Q_{1423} \Big]\op{T}{LL}{vwxy}\,.
\end{align}

\begin{table}[H]
\begin{adjustwidth}{-0.4cm}{}
\renewcommand{\arraystretch}{1.9}
\small
\begin{align*}
\begin{array}[t]{c|c|c|c}
\toprule
 \text{Operator} & \text{Tree-level Fierz} & \text{QCD shift} &  \text{QED shift}   \\
\midrule\midrule
\op{S}{LL}{q_1q_2q_3q_4}
&-\frac{1}{2}\cop{S}{LL}{q_1q_4q_3q_2}  - \frac{1}{8}\cop{T}{LL}{q_1q_4q_3q_2}
&  \makecell{-\frac{1}{8N_c}\Big[8d_1+3d_2-4d_3+10e_1-9f_1  \\+\left(32d_1-d_3-10e_2-21f_2 \right) (N_c^2-1) \Big] \op{S}{LL}{q_1q_2q_3q_4} \\ +
\frac{1}{8}\Big[8d_1 +3d_2 -4 d_3 +10 e_1 -9f_1  \Big] \cop{S}{LL}{q_1q_2q_3q_4} \\ -\frac{1}{32N_c}\Big[8d_1-d_2+10e_1+3f_1 \\
-\left(d_3+10e_2-7f_2\right)(N_c^2-1)\Big] \op{T}{LL}{q_1q_2q_3q_4} \\+ \frac{1}{32}\Big[8d_1-d_2+10e_1+3f_1  \Big] \cop{T}{LL}{q_1q_2q_3q_4}  }
& B_{q_1q_2q_3q_4}\\
&&& \\
\cop{S}{LL}{q_1q_2q_3q_4}
&-\frac{1}{2}\op{S}{LL}{q_1q_4q_3q_2}  - \frac{1}{8}\op{T}{LL}{q_1q_4q_3q_2}
& \makecell{\frac{1}{8}\Big[-32d_1+3d_2+d_3+10e_1  \\
+10e_2-9f_1+21f_2  \Big] \op{S}{LL}{q_1q_2q_3q_4} \\
- \frac{1}{8N_c}\Big[3d_2-3d_3+10e_1+10e_2-9f_1\\
+21f_2+4d_3 N_c^2-8d_1(3+N_c^2)  \Big] \cop{S}{LL}{q_1q_2q_3q_4} \\
+\frac{1}{32}\Big[-d_2+d_3+10e_1+10e_2+3f_1-7f_2 \Big] \op{T}{LL}{q_1q_2q_3q_4} \\
-\frac{1}{32N_c}\Big[8d_1-d_2+d_3+10e_1\\
+10e_2+3f_1-7f_2-8d_1N_c^2  \Big] \cop{T}{LL}{q_1q_2q_3q_4}  }
&  \widetilde B_{q_1q_2q_3q_4}\\
&&&\\
\op{S}{LL}{q_1q_2\ell_1\ell_2}
&-\frac{1}{2}\op{S}{LL}{q_1\ell_2\ell_1q_2}  - \frac{1}{8}\op{T}{LL}{q_1\ell_2\ell_1q_2}
&\makecell{\frac{1}{16N_c}\left(32d_1-d_3-10e_2-21f_2\right)(1-N_c^2)\op{S}{LL}{q_1q_2\ell_1\ell_2}\\
+ \frac{1}{64N_c}\left(d_3+10e_2-7f_2\right)(N_c^2-1)\op{T}{LL}{q_1q_2\ell_1\ell_2}}
& B_{q_1q_2\ell_1\ell_2} \\
&&&\\
\op{S}{LL}{q_1\ell_2\ell_1q_2}
&-\frac{1}{2}\op{S}{LL}{q_1q_2\ell_1\ell_2}  - \frac{1}{8}\op{T}{LL}{q_1q_2\ell_1\ell_2}
& \makecell{\frac{1}{4N_c}\left(2d_1-d_3\right)(N_c^2-1)\cop{S}{LL}{q_1\ell_2\ell_1q_2}\\
+ \frac{1}{8N_c}d_1(N_c^2-1)\cop{T}{LL}{q_1\ell_2\ell_1q_2}}
& B_{q_1\ell_2\ell_1q_2} \\
&&&\\
\op{S}{LL}{\ell_1\ell_2\ell_3\ell_4}
&-\frac{1}{2}\op{S}{LL}{\ell_1\ell_4\ell_3\ell_2}  - \frac{1}{8}\op{T}{LL}{\ell_1\ell_4\ell_3\ell_2}
& 0
&  B_{\ell_1\ell_2\ell_3\ell_4} \\
\bottomrule
\end{array}
\end{align*}
\end{adjustwidth}
\caption{Fierz transformations for scalar $P_L \otimes P_L$ operators together with their one-loop QCD and QED shifts in the generalised BMU scheme.}
\label{tab:SLLgenbmu}
\end{table}

\begin{table}[H]
\begin{adjustwidth}{-0.4cm}{}
\renewcommand{\arraystretch}{1.9}
\small
\begin{align*}
\begin{array}[t]{c|c|c|c}
\toprule
 \text{Operator} & \text{Tree-level Fierz} & \text{QCD shift} &  \text{QED shift}   \\
\midrule\midrule
\op{T}{LL}{q_1q_2q_3q_4}
&-6\cop{S}{LL}{q_1q_4q_3q_2}  + \frac{1}{2}\cop{T}{LL}{q_1q_4q_3q_2}
& \makecell{\frac{1}{2N_c}\Big[-24d_1+3d_2+50e_1+40e_2-9f_1\\
+\left(3d_3-10e_2-21f_2\right) (N_c^2-1) \Big] \op{S}{LL}{q_1q_2q_3q_4} \\
+ \frac{1}{2}\Big[24d_1-3d_2-50e_1-40e_2+9f_1  \Big] \cop{S}{LL}{q_1q_2q_3q_4} \\ +\frac{1}{8N_c}\Big[-24d_1+3d_2+10e_1-9f_1+28f_2\\
+\left(3d_3-10e_2+7f_2\right)(N_c^2-1) \Big] \op{T}{LL}{q_1q_2q_3q_4} \\
+ \frac{1}{8}\Big[24d_1-3d_2-10e_1+9f_1-28f_2 \Big] \cop{T}{LL}{q_1q_2q_3q_4}  }
& D_{q_1q_2q_3q_4} \\
&&& \\
\cop{T}{LL}{q_1q_2q_3q_4}
&-6\op{S}{LL}{q_1q_4q_3q_2}  + \frac{1}{2}\op{T}{LL}{q_1q_4q_3q_2}
& \makecell{\frac{1}{2}\Big[-3d_2+3d_3-50e_1-10e_2+9f_1-21f_2 \Big] \op{S}{LL}{q_1q_2q_3q_4} \\
+ \frac{1}{2N_c}\Big[3d_2-3d_3+50e_1+50e_2-9f_1\\
+21f_2-40e_2N_c^2+24d_1(N_c^2-1)  \Big] \cop{S}{LL}{q_1q_2q_3q_4} \\
+\frac{1}{8}\Big[-3d_2+3d_3-10e_1-10e_2+9f_1+7f_2 \Big] \op{T}{LL}{q_1q_2q_3q_4} \\
+ \frac{1}{8N_c}\Big[3d_2-3d_3+10e_1+10e_2\\
-9f_1-7f_2+4\left(6d_1-7f_2\right)(N_c^2-1)  \Big] \cop{T}{LL}{q_1q_2q_3q_4}  }
& \widetilde D_{q_1q_2q_3q_4} \\
&&& \\
\op{T}{LL}{q_1q_2\ell_1\ell_2}
&-6\op{S}{LL}{q_1\ell_2\ell_1q_2} + \frac{1}{2}\op{T}{LL}{q_1\ell_2\ell_1q_2}
& \makecell{\frac{1}{4N_c}\left[3d_3-10e_2-21f_2\right](N_c^2-1)\op{S}{LL}{q_1q_2\ell_1\ell_2}\\
+ \frac{1}{16N_c}\left[3d_3-10e_2+7f_2\right](N_c^2-1)\op{T}{LL}{q_1q_2\ell_1\ell_2}}
& D_{q_1q_2\ell_1\ell_2} \\
&&& \\
\op{T}{LL}{q_1\ell_2\ell_1q_2}
&-6\op{S}{LL}{q_1q_2\ell_1\ell_2}  + \frac{1}{2}\op{T}{LL}{q_1q_2\ell_1\ell_2}
& \makecell{\frac{2}{N_c}\left[3d_1-5e_2\right](N_c^2-1)\cop{S}{LL}{q_1\ell_2\ell_1q_2}\\+ \frac{1}{4N_c}\left[6d_1-7f_2\right](N_c^2-1)\cop{T}{LL}{q_1\ell_2\ell_1q_2}}
& D_{q_1\ell_2\ell_1q_2} \\
&&& \\
\op{T}{LL}{\ell_1\ell_2\ell_3\ell_4}
&-6\op{S}{LL}{\ell_1\ell_4\ell_3\ell_2}  + \frac{1}{2}\op{T}{LL}{\ell_1\ell_4\ell_3\ell_2}
& 0
& D_{\ell_1\ell_2\ell_3\ell_4} \\
\bottomrule
\end{array}
\end{align*}
\end{adjustwidth}
\caption{Fierz transformations for tensor $\sigma_{\mu\nu}P_L \otimes \sigma^{\mu\nu}P_L$ operators together with their one-loop QCD and QED shifts in the generalised BMU scheme.}
\label{tab:tensfullBMU}
\end{table}
\noindent
Finally we report the shifts for the TLL operators in Tab.~\ref{tab:tensfullBMU}. To abbreviate the QED shifts we use the following definition:
\begin{align}
  &D_{vwxy}\equiv \notag\\
&\frac{1}{8}\Big[\left(12d_3-40e_2-84f_2\right)Q_{1234}+\left(-12d_2-200e_1+36f_1\right)Q_{1324}\nonumber\\&\,\,\,+\left(96d_1-160e_2\right)Q_{1423} \Big]\op{S}{LL}{vwxy}\notag\\
&\frac{1}{8}\Big[\left(3d_3-10e_2+7f_2\right)Q_{1234}+\left(-3d_2-10e_1+9f_1\right)Q_{1324}\nonumber\\&\,\,\,+\left(24d_1-28f_2\right)Q_{1423} \Big]\op{T}{LL}{vwxy}\,.
\end{align}

\newpage
\newpage

\section{Poles}\label{app:1loop}
In this Appendix we report the pole structures as well as the finite terms that result from the operator insertions into the diagrams collected in App.~\ref{app:loop}. For simplicity we define the QED Wavefunction renormalization constant for an operator containing four different fermions by
\begin{equation}
  P=-\frac{1}{2}(Q_1^2+Q_2^2+Q_3^2+Q_4^2)\,.
\end{equation}
\begin{table}[H]
\begin{adjustwidth}{-0.9cm}{}
\renewcommand{\arraystretch}{1.9}
\small
\begin{align*}
\begin{array}[t]{c|c|c|c}
\toprule
 \text{Operator} & \text{WFR} \,\,[\frac{1}{\epsilon}] & \text{pole}\,\,[\frac{1}{\epsilon}] & \text{evanescent cont.}  \\
\midrule\midrule
\op{V}{LL}{q_1q_2q_3q_4} &
-2C_F\op{V}{LL}{q_1q_2q_3q_4}  &
\makecell{\left(2C_F+\frac{3}{N_c}\right)\op{V}{LL}{q_1q_2q_3q_4}\\-3\cop{V}{LL}{q_1q_2q_3q_4}} &
\makecell{\left(-4C_Fa_1+\frac{2a_3-a_2}{N_c} \right)\op{V}{LL}{q_1q_2q_3q_4}\\+\left(a_2-2a_3 \right)\cop{V}{LL}{q_1q_2q_3q_4}}\\
 & & & \\
\cop{V}{LL}{q_1q_2q_3q_4} &
-2C_F\cop{V}{LL}{q_1q_2q_3q_4}  &
\makecell{-3\op{V}{LL}{q_1q_2q_3q_4}\\+\left(2C_F+\frac{3}{N_c} \right)\cop{V}{LL}{q_1q_2q_3q_4}} &
\makecell{\left(-2a_1+a_2 \right)\op{V}{LL}{q_1q_2q_3q_4}\\+\left(-4C_Fa_3+\frac{2a_1-a_2}{N_c}\right)\cop{V}{LL}{q_1q_2q_3q_4}}\\
 & & & \\
\op{V}{LL}{q_1q_2\ell_1\ell_2} &
-C_F\op{V}{LL}{q_1q_2\ell_1\ell_2}  &
\makecell{C_F\op{V}{LL}{q_1q_2\ell_1\ell_2}} &
\makecell{-2C_Fa_1\op{V}{LL}{q_1q_2\ell_1\ell_2}}\\
 & & & \\
\op{V}{LL}{q_1\ell_1\ell_2q_2} &
-C_F\op{V}{LL}{q_1\ell_1\ell_2q_2}  &
\makecell{C_F\op{V}{LL}{q_1\ell_1\ell_2q_2}} &
\makecell{-2C_Fa_3\op{V}{LL}{q_1\ell_1\ell_2q_2}}\\
 & & & \\
 \midrule\midrule
\op{V}{LL}{\ell_1\ell_2\ell_3\ell_4} &
P\op{V}{LL}{\ell_1\ell_2\ell_3\ell_4}  &
\makecell{\left(Q_{1234}-4Q_{1324}+Q_{1423} \right)\op{V}{LL}{\ell_1\ell_2\ell_3\ell_4}} &
\makecell{\left(-2a_1Q_{1234}+a_2Q_{1324}-2a_3Q_{1423} \right)\op{V}{LL}{\ell_1\ell_2\ell_3\ell_4}}\\
\bottomrule
\end{array}
\end{align*}
\end{adjustwidth}
\caption{Loop structure for $\gamma_\mu P_L \otimes \gamma^\mu P_L$ operators: The columns correspond to the divergent WFR contributions and vertex corrections as well as the finite contributions from evanescent operators, computed in the generalised BMU scheme. The first four rows result from QCD and are given in units of $\alpha_s/(4\pi)$ whereas the last one is a QED correction given in units of $\alpha/(4\pi)$.}
\label{fig:polesVLL}
\end{table}

\begin{table}[H]
\begin{adjustwidth}{-0.9cm}{}
\renewcommand{\arraystretch}{1.9}
\small
\begin{align*}
\begin{array}[t]{c|c|c|c}
  \toprule
   \text{Operator} & \text{WFR} \,\,[\frac{1}{\epsilon}] & \text{pole}\,\,[\frac{1}{\epsilon}] & \text{evanescent cont.}  \\
  \midrule\midrule
\op{S}{LR}{q_1q_2q_3q_4} &
-2C_F\op{S}{LR}{q_1q_2q_3q_4}  &
\makecell{8C_F\op{S}{LR}{q_1q_2q_3q_4}} &
\makecell{\left(\frac{c_2+2c_3}{N_c}-8C_Fc_1 \right)\op{S}{LR}{q_1q_2q_3q_4}\\+\left(-c_2-2c_3 \right)\cop{S}{LR}{q_1q_2q_3q_4}}\\
 & & & \\
\cop{S}{LR}{q_1q_2q_3q_4} &
-2C_F\cop{S}{LR}{q_1q_2q_3q_4}  &
\makecell{3\op{S}{LR}{q_1q_2q_3q_4}\\+\left(2C_F-\frac{3}{N_c}\right)\cop{S}{LR}{q_1q_2q_3q_4}} &
\makecell{\left(-4c_1-c_2 \right)\op{S}{LR}{q_1q_2q_3q_4}\\+\left(\frac{4c_1+c_2}{N_c}-4c_3C_F \right)\cop{S}{LR}{q_1q_2q_3q_4}}\\
 & & & \\
\op{S}{LR}{q_1q_2\ell_1\ell_2} &
-C_F\op{S}{LR}{q_1q_2\ell_1\ell_2}  &
\makecell{4C_F\op{S}{LR}{q_1q_2\ell_1\ell_2}} &
\makecell{-4C_Fc_1\op{S}{LR}{q_1q_2\ell_1\ell_2}}\\
 & & & \\
\op{S}{LR}{q_1\ell_1\ell_2q_2} &
-C_F\op{S}{LR}{q_1\ell_1\ell_2q_2}  &
\makecell{C_F\op{S}{LR}{q_1\ell_1\ell_2q_2}} &
\makecell{-2C_Fc_3\op{S}{LR}{q_1\ell_1\ell_2q_2}}\\
 & & & \\
 \midrule\midrule
\op{S}{LR}{\ell_1\ell_2\ell_3\ell_4} &
P\op{S}{LR}{\ell_1\ell_2\ell_3\ell_4}  &
\makecell{\left(4Q_{1234}-Q_{1324}+Q_{1423} \right)\op{S}{LR}{\ell_1\ell_2\ell_3\ell_4}} &
\makecell{\left(-4c_1Q_{1234}-c_2Q_{1324}-2c_3Q_{1423} \right)\op{S}{LR}{\ell_1\ell_2\ell_3\ell_4}}\\
\bottomrule
\end{array}
\end{align*}
\end{adjustwidth}
\caption{Loop structure for $P_L \otimes P_R$ operators: The columns correspond to the divergent WFR contributions and vertex corrections as well as the finite contributions from evanescent operators, computed in the generalised BMU scheme. The first four rows result from QCD and are given in units of $\alpha_s/(4\pi)$ whereas the last one is a QED correction given in units of $\alpha/(4\pi)$.}
\label{fig:polesSLR}
\end{table}

\begin{table}[H]
\begin{adjustwidth}{-1cm}{}
\renewcommand{\arraystretch}{1.9}
\small
\begin{align*}
\begin{array}[t]{c|c|c|c}
  \toprule
   \text{Operator} & \text{WFR} \,\,[\frac{1}{\epsilon}] & \text{pole}\,\,[\frac{1}{\epsilon}] & \text{evanescent cont.}  \\
  \midrule\midrule
\op{S}{LL}{q_1q_2q_3q_4} &
-2C_F\op{S}{LL}{q_1q_2q_3q_4}  &
\makecell{8C_F\op{S}{LL}{q_1q_2q_3q_4} \\ -\frac{1}{2N_c}\op{T}{LL}{q_1q_2q_3q_4}  + \frac{1}{2}\cop{T}{LL}{q_1q_2q_3q_4}} &
\makecell{\left(\frac{d_3-d_2}{2N_c}-8C_Fd_1\right)\op{S}{LL}{q_1q_2q_3q_4}\\+\frac{d_2-d_3}{2}\cop{S}{LL}{q_1q_2q_3q_4}}\\
 & & & \\
\cop{S}{LL}{q_1q_2q_3q_4} &
-2C_F\cop{S}{LL}{q_1q_2q_3q_4}  &
\makecell{3\op{S}{LL}{q_1q_2q_3q_4}+\left(2C_F-\frac{3}{N_c}\right)\cop{S}{LL}{q_1q_2q_3q_4} \\+\frac{1}{4}\op{T}{LL}{q_1q_2q_3q_4}  + \left(\frac{C_F}{2}-\frac{1}{4N_c}\right)\cop{T}{LL}{q_1q_2q_3q_4}} &
\makecell{\left(\frac{d_2}{2}-4d_1\right)\op{S}{LL}{q_1q_2q_3q_4}\\+\left(\frac{4d_1}{N_c}-\frac{d_2}{2N_c}-C_Fd_3\right)\cop{S}{LL}{q_1q_2q_3q_4}} \\
 & & & \\
\op{S}{LL}{q_1q_2\ell_1\ell_2} &
-C_F\op{S}{LL}{q_1q_2\ell_1\ell_2}  &
\makecell{4C_F\op{S}{LL}{q_1q_2\ell_1\ell_2}} &
\makecell{-4C_Fd_1\op{S}{LL}{q_1q_2\ell_1\ell_2}}\\
 & & & \\
\op{S}{LL}{q_1\ell_1\ell_2q_2} &
-C_F\op{S}{LL}{q_1\ell_1\ell_2q_2}  &
\makecell{C_F\op{S}{LL}{q_1\ell_1\ell_2q_2}+\frac{C_F}{4}\op{T}{LL}{q_1\ell_1\ell_2q_2}} &
\makecell{-\frac{1}{2}C_Fd_3\op{S}{LL}{q_1\ell_1\ell_2q_2}}\\
 & & & \\
 \midrule\midrule
\op{S}{LL}{\ell_1\ell_2\ell_3\ell_4} &
P\op{S}{LL}{\ell_1\ell_2\ell_3\ell_4}  &
\makecell{\left(4Q_{1234}-Q_{1324}+Q_{1423}\right)\op{S}{LL}{\ell_1\ell_2\ell_3\ell_4}\\+\frac{1}{4}\left(Q_{1324}+Q_{1423}\right)\op{T}{LL}{\ell_1\ell_2\ell_3\ell_4}} &
\makecell{\left(-4d_1Q_{1234}+\frac{1}{2}d_2Q_{1324}-\frac{1}{2}d_3Q_{1423}\right)\op{S}{LL}{\ell_1\ell_2\ell_3\ell_4}}\\
 \bottomrule
\end{array}
\end{align*}
\end{adjustwidth}
\caption{Loop structure for $P_L \otimes P_L$ operators: The columns correspond to the divergent WFR contributions and vertex corrections as well as the finite contributions from evanescent operators, computed in the generalised BMU scheme. The first four rows result from QCD and are given in units of $\alpha_s/(4\pi)$ whereas the last one is a QED correction given in units of $\alpha/(4\pi)$.}
\label{fig:polesSLL}
\end{table}

\begin{table}[H]
\begin{adjustwidth}{-1cm}{}
\renewcommand{\arraystretch}{1.9}
\small
\begin{align*}
\begin{array}[t]{c|c|c|c}
  \toprule
   \text{Operator} & \text{WFR} \,\,[\frac{1}{\epsilon}] & \text{pole}\,\,[\frac{1}{\epsilon}] & \text{evanescent cont.}  \\
  \midrule\midrule
\op{T}{LL}{q_1q_2q_3q_4} &
-2C_F\op{T}{LL}{q_1q_2q_3q_4}  &
\makecell{-\frac{24}{N_c}\op{S}{LL}{q_1q_2q_3q_4}+24\cop{S}{LL}{q_1q_2q_3q_4}} &
\makecell{\frac{20}{N_c}(e_1+e_2)\op{S}{LL}{q_1q_2q_3q_4}-20(e_1+e_2)\cop{S}{LL}{q_1q_2q_3q_4}\\+\frac{7f_2-3f_1}{2N_c}\op{T}{LL}{q_1q_2q_3q_4}+\frac{3f_1-7f_2}{2}\cop{T}{LL}{q_1q_2q_3q_4}} \\
 & & & \\
\cop{T}{LL}{q_1q_2q_3q_4} &
-2C_F\cop{T}{LL}{q_1q_2q_3q_4}  &
\makecell{12\op{S}{LL}{q_1q_2q_3q_4}+\left(24C_F-\frac{12}{N_c}\right)\cop{S}{LL}{q_1q_2q_3q_4}\\ -3\op{T}{LL}{q_1q_2q_3q_4}  + \left(6C_F+\frac{3}{N_c}\right)\cop{T}{LL}{q_1q_2q_3q_4}} &
\makecell{-20e_1\op{S}{LL}{q_1q_2q_3q_4}+\left(-40C_Fe_2+\frac{20e_1}{N_c}\right)\cop{S}{LL}{q_1q_2q_3q_4}\\+\frac{3f_1}{2}\op{T}{LL}{q_1q_2q_3q_4}-\left(\frac{3f_1}{2N_c}+7f_2C_F\right)\cop{T}{LL}{q_1q_2q_3q_4}} \\
 & & & \\
\op{T}{LL}{q_1q_2\ell_1\ell_2} &
-C_F\op{T}{LL}{q_1q_2\ell_1\ell_2}  &
\makecell{0} &
\makecell{0}\\
 & & & \\
\op{T}{LL}{q_1\ell_1\ell_2q_2} &
-C_F\op{T}{LL}{q_1\ell_1\ell_2q_2}  &
\makecell{12C_F\op{S}{LL}{q_1\ell_1\ell_2q_2}+3C_F\op{T}{LL}{q_1\ell_1\ell_2q_2}} &
\makecell{-20C_Fe_2\op{S}{LL}{q_1\ell_1\ell_2q_2}-\frac{7}{2}C_Ff_2\op{T}{LL}{q_1\ell_1\ell_2q_2}}\\
 & & & \\
 \midrule\midrule
\op{T}{LL}{\ell_1\ell_2\ell_3\ell_4} &
P\op{T}{LL}{\ell_1\ell_2\ell_3\ell_4}  &
\makecell{\left(12Q_{1324}+12Q_{1423}\right)\op{S}{LL}{\ell_1\ell_2\ell_3\ell_4}\\+\left(-3Q_{1324}+3Q_{1423}\right)\op{T}{LL}{\ell_1\ell_2\ell_3\ell_4}} &
\makecell{\left(-20e_1Q_{1324}-20e_2Q_{1423}\right)\op{S}{LL}{\ell_1\ell_2\ell_3\ell_4}\\+\left(\frac{3}{2}f_1Q_{1324}-\frac{7}{2}f_2Q_{1423}\right)\op{T}{LL}{\ell_1\ell_2\ell_3\ell_4}}\\
\bottomrule
\end{array}
\end{align*}
\end{adjustwidth}
\caption{Loop structure for $\sigma_{\mu\nu}P_L \otimes \sigma^{\mu\nu}P_L$ operators: The columns correspond to the divergent WFR contributions and vertex corrections as well as the finite contributions from evanescent operators, computed in the generalised BMU scheme. The first four rows result from QCD and are given in units of $\alpha_s/(4\pi)$ whereas the last one is a QED correction given in units of $\alpha/(4\pi)$.}
\label{fig:polesTLL}
\end{table}

\newpage
\section{Renormalization Constants}\label{app:Zs}
In this Appendix we consider all possible operator bases for the different Dirac structures VLL, SLR, SLL and TLL and fermion combinations discussed above and augment the bases with the corresponding Fierz-evanescent operators. We report for these sets of operators the renormalisation constants, which are used in general basis transformations. In this Appendix we only consider contributions from genuine vertex corrections and neglect penguin contributions.

\subsection{Vector Operators}

\subsubsection{VLL: $q_1q_2q_3q_4$}

Starting with the most general set of VLL operators we consider the case of four different quarks:
\begin{equation}
  \{\op{V}{LL}{q_1q_2q_3q_4},\cop{V}{LL}{q_1q_2q_3q_4},\op{E}{VLL}{q_1q_2q_3q_4},\op{E}{\widetilde VLL}{q_1q_2q_3q_4}\}\,,
\end{equation}
with the evanescent operators
\begin{align}\label{eq:EVVLL}
  \op{E}{VLL}{q_1q_2q_3q_4} &\equiv \op{V}{LL}{q_1q_2q_3q_4}-\mathcal{F}\op{V}{LL}{q_1q_2q_3q_4}\,, \\
  \op{E}{\widetilde VLL}{q_1q_2q_3q_4} &\equiv \cop{V}{LL}{q_1q_2q_3q_4}-\mathcal{F}\cop{V}{LL}{q_1q_2q_3q_4}\,.
\end{align}
The renormalization constants for these operators read:
\begin{equation}
  Z_{QQ}^{(1,1)}=Z_{EE}^{(1,1)}=\left(
  \begin{array}{ccc}
    -\frac{3}{N_c} & 3  \\
    3 & -\frac{3}{N_c} \\
  \end{array} \right) \,,\quad  Z_{QE}^{(1,1)}= \left(
  \begin{array}{ccc}
    0 & 0 \\
    0 & 0 \\
  \end{array} \right) \,,
\end{equation}
and the finite renormalization is given by
\begin{equation}
  Z_{EQ}^{(1,0)}= \left(
  \begin{array}{ccc}
    2(a_1-a_3)N_c & -2 (a_1-
   a_3) \\
 2 (a_1- a_3) & -2(a_1-a_3)N_c \\
  \end{array} \right)\,.
\end{equation}

\subsubsection{VLL: $q_1q_2q_1q_2$}
For operators generating $\Delta F=2$ transitions we define the operator basis

\begin{equation}
  \{\op{V}{LL}{q_1q_2q_1q_2},\op{E}{VLL}{q_1q_2q_1q_2}\}\,,
\end{equation}
where $\op{E}{VLL}{q_1q_2q_1q_2}$ is defined analogously as in eq.~\eqref{eq:EVVLL}. One finds for the renormalization constants:

\begin{equation}
  Z_{QQ}^{(1,1)}=3-\frac{3}{N_c} \,,\quad Z_{QE}^{(1,1)}= -3 \,,\quad  Z_{EQ}^{(1,0)}= 2(a_1 - a_3) (N_c-1) \,,\quad  Z_{EE}^{(1,1)}= -3-\frac{3}{N_c} \,.
\end{equation}
The same relations hold if the basis $\{\cop{V}{LL}{q_1q_2q_1q_2},\op{E}{\widetilde VLL}{q_1q_2q_1q_2}\}$ is used instead.

\subsubsection{VLL: $q_1q_2 \ell_1\ell_2$}
For semi-leptonic operators we define the operator basis

\begin{equation}
  \{\op{V}{LL}{q_1q_2 \ell_1\ell_2},\op{E}{VLL}{q_1q_2 \ell_1\ell_2}\}\,,
\end{equation}
where $\op{E}{VLL}{q_1q_2 \ell_1\ell_2}$ is defined analogously as in eq.~\eqref{eq:EVVLL}. One finds for the renormalization constants:

\begin{equation}
  Z_{QQ}^{(1,1)}=0 \,,\quad Z_{QE}^{(1,1)}= 0 \,,\quad  Z_{EQ}^{(1,0)}= 2(a_1-a_3)C_F \,,\quad  Z_{EE}^{(1,1)}= 0 \,.
\end{equation}
The same relations hold if the basis $\{\op{V}{LL}{q_1\ell_2 \ell_1q_2},\op{E}{VLL}{q_1\ell_2 \ell_1q_2}\}$ is used instead with the only difference that $Z_{EQ}^{(1,0)}$ changes sign.

\subsubsection{VLL: $\ell_1\ell_2\ell_3\ell_4$}
For four-lepton operators we define the operator basis

\begin{equation}
  \{\op{V}{LL}{\ell_1\ell_2\ell_3\ell_4},\op{E}{VLL}{\ell_1\ell_2\ell_3\ell_4}\}\,,
\end{equation}
where $\op{E}{VLL}{\ell_1\ell_2\ell_3\ell_4}$ is defined analogously as in eq.~\eqref{eq:EVVLL}. One finds for the QED renormalization constants:

\begin{align}
  Z_{QQ}^{(1,1)}&=-P-Q_{1234}+4Q_{1324}-Q_{1423} \,,\\
   Z_{QE}^{(1,1)}&= 0 \,, \\
   Z_{EQ}^{(1,0)}&= 2 (a_1 - a_3) (Q_1 - Q_3) (Q_2 - Q_4) \,,\\
   Z_{EE}^{(1,1)}&= Z_{QQ}^{(1,1)}\,.
\end{align}

\subsection{Scalar and Vector LR Operators}
\subsubsection{SLR: $q_1q_2q_3q_4$}

Starting with the most general set of SLR operators we consider the case of four different quarks:

\begin{equation}
  \{\op{S}{LR}{q_1q_2q_3q_4},\cop{S}{LR}{q_1q_2q_3q_4},\op{E}{SLR}{q_1q_2q_3q_4},\op{E}{\widetilde SLR}{q_1q_2q_3q_4}\}\,,
\end{equation}
with the evanescent operators
\begin{align}\label{eq:EVSLR}
  \op{E}{SLR}{q_1q_2q_3q_4} &\equiv \op{S}{LR}{q_1q_2q_3q_4}-\mathcal{F}\op{S}{LR}{q_1q_2q_3q_4}\,, \\
  \op{E}{\widetilde SLR}{q_1q_2q_3q_4} &\equiv \cop{S}{LR}{q_1q_2q_3q_4}-\mathcal{F}\cop{S}{LR}{q_1q_2q_3q_4}\,.
\end{align}
The renormalization constants for these operators read:
\begin{equation}
  Z_{QQ}^{(1,1)}=Z_{EE}^{(1,1)}=\left(
  \begin{array}{ccc}
    -6C_F & 0  \\
    -3 & \frac{3}{N_c} \\
  \end{array} \right) \,,\quad  Z_{QE}^{(1,1)}= \left(
  \begin{array}{ccc}
    0 & 0 \\
    0 & 0 \\
  \end{array} \right) \,,
\end{equation}
and the finite renormalization is given by
\begin{equation}
  Z_{EQ}^{(1,0)}= \left(
  \begin{array}{ccc}
    \frac{b_2 + 2b_1- 2 c_3 - c_2}{N_c}-8 b_3 C_F  + 8 c_1 C_F & c_2+2c_3-2b_1-b_2 \\
 -b_2 - 4 b_3 + 4 c_1 + c_2 & \frac{b_2 + 4 b_3 - 4 c_1 - c_2}{N_c}-4 b_1 C_F  + 4 c_3 C_F \\
  \end{array} \right)\,.
\end{equation}

\subsubsection{SLR: $q_1q_2q_1q_2$}
For operators generating $\Delta F=2$ transitions we choose to eliminate the Parity-flipped operators $\op{S}{RL}{q_1q_2q_1q_2},\cop{S}{RL}{q_1q_2q_1q_2}$ and therefore the initial basis together with the renormalization constants stays the same as in the previous subsection. The basis takes the form

\begin{equation}
  \{\op{S}{LR}{q_1q_2q_1q_2},\cop{S}{LR}{q_1q_2q_1q_2},\op{E}{SLR}{q_1q_2q_1q_2},\op{E}{\widetilde SLR}{q_1q_2q_1q_2}\}\,.
\end{equation}

\subsubsection{SLR: $q_1q_2 \ell_1\ell_2$}
For semi-leptonic operators we define the operator basis

\begin{equation}
  \{\op{S}{LR}{q_1q_2 \ell_1\ell_2},\op{E}{SLR}{q_1q_2 \ell_1\ell_2}\}\,,
\end{equation}
where $\op{E}{SLR}{q_1q_2 \ell_1\ell_2}$ is defined analogously as in eq.~\eqref{eq:EVSLR}. One finds for the renormalization constants:

\begin{equation}
  Z_{QQ}^{(1,1)}=-3C_F \,,\quad Z_{QE}^{(1,1)}= 0 \,,\quad  Z_{EQ}^{(1,0)}= 4(c_1-b_3)C_F \,,\quad  Z_{EE}^{(1,1)}= -3C_F \,.
\end{equation}
For the basis $\{\op{S}{LL}{q_1\ell_2 \ell_1q_2},\op{E}{SLL}{q_1\ell_2 \ell_1q_2}\}$ one finds the following renormalization constants:
\begin{equation}
  Z_{QQ}^{(1,1)}=0 \,,\quad Z_{QE}^{(1,1)}= 0 \,,\quad  Z_{EQ}^{(1,0)}= 2 (c_3-b_1) C_F \,,\quad  Z_{EE}^{(1,1)}= 0 \,.
\end{equation}

\subsubsection{SLR: $\ell_1\ell_2\ell_3\ell_4$}
For four-lepton operators we define the operator basis
\begin{equation}
  \{\op{S}{LR}{\ell_1\ell_2\ell_3\ell_4},\op{E}{SLR}{\ell_1\ell_2\ell_3\ell_4}\}\,,
\end{equation}
where $\op{E}{VLL}{\ell_1\ell_2\ell_3\ell_4}$ is defined analogously as in eq.~\eqref{eq:EVSLR}. One finds for the QED renormalization constants:

\begin{align}
  Z_{QQ}^{(1,1)}&=-P-4Q_{1234}+Q_{1324}-Q_{1423} \,,\\
   Z_{QE}^{(1,1)}&= 0 \,, \\
   Z_{EQ}^{(1,0)}&= 2 Q_{1423} (c_3-b_1)+Q_{1324}(c_2-b_2)+4Q_{1234} (c_1-b_3) \,,\\
   Z_{EE}^{(1,1)}&= Z_{QQ}^{(1,1)}\,.
\end{align}

\subsection{Scalar and Tensor LL Operators}

\subsubsection{SLL, TLL: $q_1q_2q_3q_4$}

Starting with the most general set of SLL and TLL operators we consider the case of four different quarks:
\begin{equation}
  \{\op{S}{LL}{q_1q_2q_3q_4},\cop{S}{LL}{q_1q_2q_3q_4},\op{T}{LL}{q_1q_2q_3q_4},\cop{T}{LL}{q_1q_2q_3q_4},\op{E}{SLL}{q_1q_2q_3q_4},\op{E}{\widetilde SLL}{q_1q_2q_3q_4},\op{E}{TLL}{q_1q_2q_3q_4},\op{E}{\widetilde TLL}{q_1q_2q_3q_4}\}\,,
\end{equation}
with the evanescent operators

\begin{align}\label{eq:EVSLL}
  \op{E}{SLL}{q_1q_2q_3q_4} &\equiv \op{S}{LL}{q_1q_2q_3q_4}-\mathcal{F}\op{S}{LL}{q_1q_2q_3q_4}\,, \\
  \op{E}{\widetilde SLL}{q_1q_2q_3q_4} &\equiv \cop{S}{LL}{q_1q_2q_3q_4}-\mathcal{F}\cop{S}{LL}{q_1q_2q_3q_4}\,,\\
  \op{E}{TLL}{q_1q_2q_3q_4} &\equiv \op{T}{LL}{q_1q_2q_3q_4}-\mathcal{F}\op{T}{LL}{q_1q_2q_3q_4}\,, \\
  \op{E}{\widetilde TLL}{q_1q_2q_3q_4} &\equiv \cop{T}{LL}{q_1q_2q_3q_4}-\mathcal{F}\cop{T}{LL}{q_1q_2q_3q_4}\,.
\end{align}
The renormalization constants for these operators read:

\begin{equation}
  Z_{QQ}^{(1,1)}=Z_{EE}^{(1,1)}=\left(
  \begin{array}{cccc}
    -6 C_F & 0 &
     \frac{1}{2 N_c} & -\frac{1}{2} \\
   -3 & \frac{3}{N_c} &
     -\frac{1}{4} & -\frac{N_c^2-2}{4 N_c} \\
   \frac{24}{N_c} & -24 & 2 C_F & 0 \\
   -12 & -\frac{12 \left(N_c^2-2\right)}{N_c} & 3
     & 2 C_F-3 N_c \\
  \end{array} \right) \,,
\end{equation}
and the finite renormalization we find:
\begin{align}
&\left(Z_{EQ}^{(1,0)}\right)_{11}= -\frac{-32 d_1 N_c^2+24 d_1-3
      d_2+d_3 N_c^2+3 d_3-10
      e_1+10 e_2 N_c^2-10 e_2}{8
      N_c}\\ & \quad \qquad \qquad \,\,\,\,-\frac{9
      f_1+21 f_2 N_c^2-21 f_2}{8
      N_c}\,,\nonumber\\
&\left(Z_{EQ}^{(1,0)}\right)_{12}= \frac{1}{8} (-8 d_1-3 d_2+4
      d_3-10 e_1+9 f_1)\,,\\
& \left(Z_{EQ}^{(1,0)}\right)_{13}=\frac{8
      d_1-d_2-d_3 N_c^2+d_3+10
      e_1-10 e_2 N_c^2+10 e_2+3
      f_1}{32
      N_c}\\ & \quad \qquad \qquad \,\,\,\,+\frac{7 f_2 N_c^2-7 f_2}{32
      N_c} \,, \nonumber\\
&\left(Z_{EQ}^{(1,0)}\right)_{14}= \frac{1}{32} (-8 d_1+d_2-10
      e_1-3 f_1) \,,\\
&\left(Z_{EQ}^{(1,0)}\right)_{21}=\frac{1}{8} (32 d_1-3 d_2-d_3-10
      e_1-10 e_2+9 f_1-21 f_2) \,,\\
&\left(Z_{EQ}^{(1,0)}\right)_{22}= \frac{-8 d_1 \left(N_c^2+3\right)+3
      d_2+4 d_3 N_c^2-3 d_3+10
      e_1+10 e_2}{8
      N_c}\\ & \quad \qquad \qquad+\frac{-9 f_1+21 f_2}{8
      N_c}\,, \nonumber\\
&\left(Z_{EQ}^{(1,0)}\right)_{23}=\frac{1}{32} (d_2-d_3-10
      e_1-10 e_2-3 f_1+7 f_2)\,, \\
&\left(Z_{EQ}^{(1,0)}\right)_{24}=\frac{-8 d_1 N_c^2+8
      d_1-d_2+d_3+10 e_1+10 e_2+3
      f_1-7 f_2}{32 N_c} \,,\\
&\left(Z_{EQ}^{(1,0)}\right)_{31}=\frac{24 d_1-3 d_2-3 d_3 N_c^2+3
      d_3-50 e_1+10 e_2 N_c^2-50
      e_2}{2 N_c}\\ & \quad \qquad \qquad+\frac{9 f_1+21 f_2 N_c^2-21
      f_2}{2 N_c} \,,\nonumber\\
&\left(Z_{EQ}^{(1,0)}\right)_{32}=\frac{1}{2} (-24 d_1+3
      d_2+50 e_1+40 e_2-9 f_1)\,,\\
&\left(Z_{EQ}^{(1,0)}\right)_{33}=-\frac{-24 d_1+3 d_2+3 d_3
      N_c^2-3 d_3+10 e_1-10 e_2
      N_c^2+10 e_2-9 f_1}{8 N_c}\\ & \quad \qquad \qquad \,\,\,\,-\frac{7 f_2
      N_c^2+21 f_2}{8 N_c} \,, \nonumber\\
&\left(Z_{EQ}^{(1,0)}\right)_{34}=\frac{1}{8} (-24
      d_1+3 d_2+10 e_1-9 f_1+28
      f_2)\,,\\
&\left(Z_{EQ}^{(1,0)}\right)_{41}=\frac{1}{2} (3 d_2-3 d_3+50 e_1+10
      e_2-9 f_1+21 f_2) \,,\\
&\left(Z_{EQ}^{(1,0)}\right)_{42}=-\frac{24 d_1
      \left(N_c^2-1\right)+3 d_2-3 d_3+50
      e_1-40 e_2 N_c^2+50 e_2}{2 N_c}\\ & \quad \qquad \qquad \,\,\,\,-\frac{-9
      f_1+21 f_2}{2 N_c}\,, \nonumber\\
&\left(Z_{EQ}^{(1,0)}\right)_{43}=\frac{1}{8} (3
      d_2-3 d_3+10 e_1+10 e_2-9
      f_1-7 f_2)\,, \\
&\left(Z_{EQ}^{(1,0)}\right)_{44}=-\frac{24 d_1
      \left(N_c^2-1\right)+3 d_2-3 d_3+10
      e_1+10 e_2-9 f_1}{8 N_c}\\ & \quad \qquad \qquad \,\,\,\,-\frac{-28 f_2
      N_c^2+21 f_2}{8 N_c}\,. \nonumber
\end{align}

\subsubsection{SLL, TLL: $q_1q_2q_1q_2$}
For operators generating $\Delta F=2$ transitions we choose the basis which is closely related to the one chosen in \cite{Gorbahn:2009pp}, namely

\begin{equation}
  \{\op{S}{LL}{q_1q_2q_1q_2},\op{T}{LL}{q_1q_2q_1q_2},\op{E}{\widetilde SLL}{q_1q_2q_1q_2},\op{E}{\widetilde TLL}{q_1q_2q_1q_2}\}\,,
\end{equation}
where the two evanescent operators are defined according to eq.~\eqref{eq:EVSLL}. We find for the renormalization constants:

\begin{equation}
  Z_{QQ}^{(1,1)}=\left(
  \begin{array}{ccc}
    3-6C_F & \frac{2-N_c}{4
     N_c} \\
   \frac{12
     (N_c+2)}{N_c} &
     2 C_F+3 \\
  \end{array} \right) \,,\quad  Z_{QE}^{(1,1)}= \left(
  \begin{array}{ccc}
    0 & -\frac{1}{2} \\
    -24 & 0 \\
  \end{array} \right) \,,
\end{equation}
and the finite renormalization reads:
\begin{align}
&\left(Z_{EQ}^{(1,0)}\right)_{11}=\frac{N_c (16 d_1
       (N_c+2)-3 d_2-2
       d_3
       N_c-d_3+9
       f_1-21 f_2)-10
       e_1 (N_c+2)}{8
       N_c}\\ & \quad \qquad \qquad-\frac{10
       e_2 (N_c+2)}{8
       N_c}\,, \nonumber\\
&\left(Z_{EQ}^{(1,0)}\right)_{12}= \frac{16
       d_1+d_2
       (N_c-2)-N_c (2
       d_3
       N_c+d_3+10
       e_1+10 e_2+3
       f_1-7 f_2)}{32 N_c}\\ & \quad \qquad \qquad+\frac{2
       (d_3+3 f_1-7
       f_2)}{32 N_c}\,,\nonumber\\
&\left(Z_{EQ}^{(1,0)}\right)_{21}=\frac{N_c (-3
       d_3+50 e_1-20
       e_2 N_c+10
       e_2-9 f_1-42
       f_2 N_c+21
       f_2)}{2
       N_c} \\ & \quad \qquad \qquad +\frac{48 d_1
       \left(N_c^2-1\right)+3
       d_2
       (N_c+2) -6 d_3+40
       e_1+40 e_2-18
       f_1+42 f_2}{2
       N_c}\,, \nonumber\\
&\left(Z_{EQ}^{(1,0)}\right)_{22}=\frac{N_c
       (3 d_2-3 d_3-9
       f_1+14 f_2
       N_c-7 f_2)+10
       e_1 (N_c+2)}{8 N_c}\\ & \quad \qquad \qquad +\frac{10
       e_2 \left(-2
       N_c^2+N_c+2\right)}{8 N_c}\,, \nonumber
\end{align}

\begin{equation}
  Z_{EE}^{(1,1)}= \left(
  \begin{array}{ccc}
    \begin{array}{cc}
      -3+\frac{3}{N_c} & \frac{1}{4}
    \left(-N_c+\frac{2}{N_c}-1\right) \\
  -12
    N_c+\frac{24}{N_c}+12 & 2 C_F-3
    (N_c+1) \\
    \end{array}
  \end{array} \right)\,.
\end{equation}

\subsubsection{SLL, TLL: semi-leptonic}
\paragraph{SLL and TLL ($q_1q_2 \ell_1\ell_2$):} For semi-leptonic operators we define the operator basis
\begin{equation}
  \{\op{S}{LL}{q_1q_2 \ell_1\ell_2},\op{T}{LL}{q_1q_2 \ell_1\ell_2},\op{E}{ SLL}{q_1q_2 \ell_1\ell_2},\op{E}{ TLL}{q_1q_2 \ell_1\ell_2}\}\,,
\label{basennf}
\end{equation}
where the two evanescent operators are defined according to eq.~\eqref{eq:EVSLL}. One finds for the renormalization constants:
\begin{equation}
  Z_{QQ}^{(1,1)}=Z_{EE}^{(1,1)}=\left(
  \begin{array}{ccc}
    -3C_F & 0 \\
   0 &
     C_F \\
  \end{array} \right)\,, \quad  Z_{QE}^{(1,1)}=\left(
  \begin{array}{ccc}
    0 & 0 \\
   0 &
     0 \\
  \end{array} \right)\, ,
\end{equation}

\begin{equation}
  Z_{EQ}^{(1,0)}=\left(
  \begin{array}{ccc}
   \frac{C_F}{8}\left(32d_1-d_3-10e_2-21f_2 \right) &
     \frac{C_F}{32}\left(-d_3-10e_2+7f_2 \right) \\
   \frac{C_F}{2}\left(-3d_3+10e_2+21f_2 \right) &
     \frac{C_F}{8}\left(-3d_3+10e_2-7f_2 \right) \\
  \end{array} \right) \, .
\end{equation}

\paragraph{SLL and TLL ($q_1\ell_2 \ell_1 q_2$):} Similarly, we can choose to work in the fierzed basis with respect to eq.~\eqref{basennf} :
\begin{equation}
  \{\op{S}{LL}{q_1 \ell_2\ell_1q_2},\op{T}{LL}{q_1 \ell_2\ell_1q_2},\op{E}{  SLL}{q_1 \ell_2\ell_1q_2},\op{E}{ TLL}{q_1 \ell_2\ell_1q_2}\}\,,
\end{equation}
where the two evanescent operators are defined according to eq.~\eqref{eq:EVSLL}. One finds for the renormalization constants:
\begin{equation}
  Z_{QQ}^{(1,1)}=Z_{EE}^{(1,1)}=\left(
  \begin{array}{ccc}
    0 & -\frac{C_F}{4} \\
   -12C_F &
     -2C_F \\
  \end{array} \right)\,, \quad  Z_{QE}^{(1,1)}=\left(
  \begin{array}{ccc}
    0 & 0 \\
   0 &
     0 \\
  \end{array} \right)\, ,
\end{equation}

\begin{equation}
  Z_{EQ}^{(1,0)}=\left(
  \begin{array}{ccc}
   \frac{C_F}{2}\left(d_3-2d_1 \right) &
     -\frac{C_F}{4}d_1 \\
   C_F\left(20e_2-12d_1 \right) &
     \frac{C_F}{2}\left(7f_2-6d_1 \right) \\
  \end{array} \right) \, .
\end{equation}

\subsubsection{SLL, TLL: $\ell_1\ell_2\ell_3\ell_4$}
For four-lepton operators we define the operator basis
\begin{equation}
  \{\op{S}{LL}{\ell_1\ell_2\ell_3\ell_4},\op{T}{LL}{\ell_1\ell_2\ell_3\ell_4},\op{E}{SLL}{\ell_1\ell_2\ell_3\ell_4},\op{E}{TLL}{\ell_1\ell_2\ell_3\ell_4}\}\,,
\end{equation}
where the two evanescent operators are defined according to eq.~\eqref{eq:EVSLL}. One finds for the QED renormalization constants:

\begin{equation}
  Z_{QQ}^{(1,1)}=Z_{EE}^{(1,1)}=\left(
  \begin{array}{ccc}
    -P+Q_{1324}-4Q_{1234}-Q_{1423} & -\frac{1}{4}\left(Q_{1324}+Q_{1423}\right) \\
   -12Q_{1324}-12Q_{1423} &
     -P+3Q_{1324}-3Q_{1423} \\
  \end{array} \right)\,,
\end{equation}

\begin{equation}
  Z_{QE}^{(1,1)}=\left(
  \begin{array}{ccc}
    0 & 0 \\
   0 &
     0 \\
  \end{array} \right)\,,
\end{equation}
\newpage
\begin{align}
&\left(Z_{EQ}^{(1,0)}\right)_{11}=\frac{1}{8}\big[\left(32d_1-d_3-10e_2-21f_2\right)Q_{1234}+\left(-3d_2-10e_1+9f_1\right)Q_{1324}\\&\qquad \qquad \quad \,\,\,\,+\left(4d_3-8d_1\right)Q_{1423} \big] \,,\nonumber\\
&\left(Z_{EQ}^{(1,0)}\right)_{12}= \frac{1}{32}\big[\left(-d_3-10e_2+7f_2\right)Q_{1234}+\left(d_2-10e_1-3f_1\right)Q_{1324}\\&\qquad \qquad \quad \,\,\,\,-\left(8d_1\right)Q_{1423} \big]\,,\nonumber\\
&\left(Z_{EQ}^{(1,0)}\right)_{21}=\frac{1}{2}\big[\left(-3d_3+10e_2+21f_2\right)Q_{1234}+\left(3d_2+50e_1-9f_1\right)Q_{1324}\\&\qquad \qquad \quad \,\,\,\,+\left(-24d_1+40e_2\right)Q_{1423} \big]\,, \nonumber\\
&\left(Z_{EQ}^{(1,0)}\right)_{22}=\frac{1}{8}\big[\left(-3d_3+10e_2-7f_2\right)Q_{1234}+\left(3d_2+10e_1-9f_1\right)Q_{1324}\\&\qquad \qquad \quad \,\,\,\,+\left(28f_2-24d_1\right)Q_{1423} \big]\,. \nonumber
\end{align}

\bibliographystyle{JHEP}

\bibliography{refs}

\end{document}